%% file: preprint.tex
\documentclass[letterpaper,times]{IONconf_pre}

\usepackage{wrapfig,booktabs,fancyhdr,amsmath,amsfonts,tabularx,numprint}
\usepackage{bm,bbm,amssymb,amsthm,url,multirow,times,enumitem,comment}
\usepackage{mathtools,siunitx,balance,tikz,adjustbox,graphicx,array, dsfont}
\usepackage{subcaption}
\usepackage[linesnumbered,ruled]{algorithm2e}
\usepackage[hidelinks]{hyperref}
\usepackage{float}

\usepackage{natbib}
\newcommand{\vb}{\boldsymbol}

\newcommand{\vbb}[1]{\bar{\boldsymbol{#1}}}
\newcommand{\vbt}[1]{\tilde{\boldsymbol{#1}}}

\newcommand{\T}{^{\mathsf{T}}}

\DeclareMathOperator*{\argmin}{\arg\!\min}

\newif\ifpreprint
\preprinttrue

\begin{document}

\setlength\parindent{0pt}
\setlength\parskip{6pt}

\title{Chasing Lightning: Detecting, Characterizing, and Identifying a Powerful
  Space-Based GNSS Interference Source } \author{Zachary L. Clements$^*$,
  Argyris Kriezis\textsuperscript{\textdagger}, and Todd E. Humphreys$^*$
  \vspace{1mm}
  \\
  \vspace{1mm}
  \emph{ $^*$ Department of Aerospace Engineering and Engineering Mechanics, The University of Texas at Austin} \\
  \emph{ \textsuperscript{\textdagger} Department of Aeronautics and
    Astronautics, Stanford University} \\
}

\maketitle

\pagestyle{plain}

\section*{Abstract}
\input{abstract}

\input{paper_common}

\bibliographystyle{apalike}
\bibliography{local}
\end{document}

%% file: abstract.tex
This paper analyzes and identifies a space-based Global Navigation Satellite
System (GNSS) interference source that has caused scores of powerful transient
wide-area interference events over continental Europe, Greenland, and Canada
since 2019.  While terrestrial or near-terrestrial sources are primarily
responsible for the recent uptick in GNSS interference worldwide, space-based
interferers are of special concern given their potential for vast geographic
reach and their portent of a qualitative escalation in GNSS interference.  Based
on data collected between 2019 and 2026 from a network of terrestrial GNSS
reference stations, this paper (1) develops a received-power-based detection
framework; (2) details the spatial, temporal, and spectral patterns of wide-area
interference events caused by the source; (3) presents and analyzes
identification techniques that blend received-power and
time-difference-of-arrival measurements; and (4) applies these techniques to
confidently identify the GNSS interference source as a constellation of Russian
early warning satellites in Molniya (``lightning'') orbits.

%% file: paper_common.tex
\section{Introduction}
Global Navigation Satellite Systems (GNSS) such as GPS provide meter-accurate
positioning while offering global accessibility and all-weather, radio-silent
operation.  However, GNSS is fragile: its service is easily degraded by both
deliberate (e.g., jamming and spoofing) and naturally occurring (e.g., multipath
and atmospheric) interference \citep{scott2003asa, t_humphreys_gcs08,
  humphreys2012congressionaltestimony, humphreysgnsshandbook}.  The past five
years have seen a crescendo of GNSS disruptions in the aviation and maritime
sectors \citep{osechas2022impact, opsgroup2024spoofing, perezmarcos2018,
  rin2026spoofing}.  Interference is especially problematic in these fields
because GNSS is deeply interwoven into critical systems.  Beyond simple
navigational or timing displacement, cascading failures can occur when GNSS loss
or corruption triggers follow-on problems in downstream systems
\citep{opsgroup2024spoofing, rin2026spoofing}.

Fortunately, significant progress has been made over the past decades in GNSS
interference mitigation and countermeasures \citep{psiaki2016gnssspoofing,
  psiakinewbluebookspoofing, borio2012gnss, clements2022cpimuspoofionitm}.  It
has been shown that dedicated GNSS interference monitoring systems provide
valuable spectrum situational awareness to enhance navigation security
\citep{gunawardena2009anomalous, gunawardena2011multi, stader2021leveraging,
  kriezis2024gnss, sokolova2022characterization, Morrisonnavi.560}, with some
demonstrating the ability to geolocate the interference source
\citep{bhatti2012devanddemo,mitch2016chirp}.  Furthermore, GNSS receivers
situated in LEO enable terrestrial GNSS interference detection,
characterization, and geolocation with worldwide coverage \citep{murrian2021leo,
  clements2023plansdirectgeo, clements2026spoofergeo}.

Networks of terrestrial GNSS receivers can also be enlisted for GNSS
interference monitoring \citep{abraha2024gnss}, including of space-based
interference.  \citet{york2014detailed} revealed that GPS PRNs 24 and 27 (SVN65
and SVN66) were transmitting leakage tones at $\pm$10 MHz and $\pm$20 MHz from
the GPS L5 frequency.  The anomaly was subsequently fixed.  More recently, a
network of terrestrial reference stations was used to identify a BeiDou
satellite (NORAD ID 40749) as a source of interference in the B3I band (1268.52
MHz) \citep{patil2023detecting, patil2024detecting}.  The offending satellite
continuously transmitted a tone centered at 1268.52 MHz and tones at $\pm$10 MHz
and $\pm$20 MHz from this frequency.  These tones are no longer present at the
time of writing.  Both of these interference cases were apparently due to faulty
or improperly configured hardware.

This paper reports on the detection, characterization, and source identification
of powerful transient GNSS interference originating from a non-GNSS
satellite. The phenomenon was briefly mentioned in \citet{cginavisp}, and more
fully characterized in a preliminary conference publication by the current
authors, \citet{clements2025spaceintgnss}.  The current paper significantly
extends both works.  The interference reported here is different from that
detailed in \citet{patil2023detecting, patil2024detecting, york2014detailed} in
several respects: (1) it is transient rather than continuous---the duration of
interference events is less than 10 seconds; (2) it affects signals in the
widely used GPS L1 frequency band rather than the less-common L5 and B3I bands;
(3) it is much more powerful, causing drops in terrestrial receiver
carrier-to-noise ratios (CNRs) of up to 10 dB; and (4) it is not merely an
anomaly of existing GNSS signals, nor does it originate from a GNSS satellite.

Obviously, high-powered interference with continental reach affecting the GPS L1
band---the primary band used for global aviation, shipping, and precise
timing---is of serious concern.  If deliberate, it portends a qualitative
escalation in GNSS interference.

The effects of this interference are evident in public data from a network of
terrestrial reference stations operated by the International GNSS Service (IGS)
\citep{dow2009international, kouba2009guide, montenbruck2017multi,
  johnstongnsshandbook}.  Most useful are data from a subset of IGS stations
that produce high-rate (1-Hz) GNSS observables.  These data may be retrieved
from the Crustal Dynamics Data Information System archive, which is made
available through NASA's archive of space geodesy data \citep{noll2010crustal}.
On scores of occasions since 2019, all tracked signals on the GPS L1 frequency
at IGS reference stations across Europe, Greenland, and Canada simultaneously
saw a sudden brief drop in CNR.  The onset of the disruption was synchronous to
within the 1-Hz sampling resolution, suggesting a single source per event.  The
affected terrestrial receivers span a geographic area so large that no single
ground-based or aircraft-based source could reach them all; hence the
space-based origin hypothesis.

While CNR variations across a receiver network are a valuable metric for
space-based interference detection, it will be shown that they are not
sufficiently informative for unique source identification.  By contrast, a
source's position and velocity can be instantaneously estimated using time- and
frequency-difference-of-arrival (T/FDOA) techniques if a spatially diverse
network of four or more ground stations simultaneously captures raw broadband
samples in the band affected by the interference.  Moreover, by referencing a
catalog of satellite ephemerides (assuming the source satellite is listed), even
two stations are adequate to narrow the possibilities to a manageable few.  An
association framework that fuses both CNR and TDOA measurements can then allow a
unique identification, as will be shown.

This paper makes six main contributions.  First, it presents measurement models
and a detection framework for transient wide-area interference. Second, it
details the spatial, temporal, and spectral properties of multiple wide-area
GNSS outage events from the space-based interference source and distinguishes
these from a naturally occurring solar radio burst.  Third, it presents a basic
satellite identification strategy to narrow the candidate satellites and
estimate the minimum satellite altitude at apogee.  Fourth, it presents an
advanced satellite associate framework using the generalized likelihood ratio
test (GRLT) and applies the framework to a test scenario. Fifth, it presents a
framework for instantaneously identifying an interference satellite based on a
brief time history of TDOA measurements, and gives an error sensitivity
analysis.  Sixth, it combines IGS CNR data and raw wideband samples from two
additional receivers in Europe to confidently identify the source, which is
revealed to be a small constellation of Russian satellites in Molniya
(``lightning'') orbits.

Compared to the preliminary conference version of this paper in
\citet{clements2025spaceintgnss}, contributions two and three are extended and
contributions four, five, and six are novel.

\section{Measurement Models and Detection}

The IGS reference station network collects and provides observables in the
Receiver Independent Exchange Format (RINEX) \citep{johnstongnsshandbook}.
These GPS-time-tagged observables include carrier phase, pseudorange, Doppler,
and CNR measurements for each tracked GNSS satellite. CNR observables from GPS
L1 C/A signals at stations providing high-rate (1-Hz) GNSS observables are the
focus of the following analysis.

\subsection{Measurement Models}
Let $\mathcal{I}$ be the set of all relevant terrestrial reference stations
(primarily IGS stations), and let $\mathcal{J}$ be the set of all unique GNSS
signals that can be tracked by such stations at a given center frequency (e.g.,
the GPS L1 frequency).  By way of notation, let $|\mathcal{I}|$ denote the
cardinality of $\mathcal{I}$, and similarly for all other sets. For station
$i \in \mathcal{I}$ and signal $j \in \mathcal{J}$, let CNR$_{ij}$ denote the
true CNR, expressed as
\begin{align}
  \label{eq:CNRij1}
  \text{CNR}_{ij} &= P^{ij}_\text{R} - N^i_0\quad\quad\text{(dB-Hz)}\\
  \label{eq:PijR}
  P^{ij}_\text{R} &= P^j_\text{T} + G^j_\text{T}\left(\theta^{ij}_\text{T},  \phi^{ij}_\text{T}\right) + G_\text{R}^i\left(\theta^{ij}_\text{R},  \phi^{ij}_\text{R}\right) + L_{ij} \quad\quad\text{(dBW)}
\end{align}
where $P^{ij}_\text{R}$ is the received power (dBW), $N^i_0$ is the
thermal noise density (dBW/Hz), $P^j_\text{T}$ is the transmit power (dBW),
$G^j_\text{T}$ is the transmitter antenna's gain function (dB), $G_\text{R}^i$ is
the receiver antenna's gain function (dB), $\theta^{ij}_\text{R}$ and
$\theta^{ij}_\text{T}$ are the off-boresight angles at the receiver and
transmitter, $\phi^{ij}_\text{R}$ and $\phi^{ij}_\text{T}$ are the azimuth
angles at the receiver and transmitter, and $L_{ij}$ is the path loss (dB),
defined as
\begin{align}
  \label{eq:LijPathLoss}
  L_{ij} &= 20\,\text{log}_{10} \left( \frac{\lambda}{4\pi\rho_{ij}}\right) \quad\quad\text{(dB)}
\end{align}
where $\lambda$ is the carrier wavelength and $\rho_{ij}$ is the range to the
satellite transmitting signal $j$. The noise power density $N^i_0$ also includes
the effects of multi-access interference from GNSS signals other than the $j$th
one on the same frequency.

CNR becomes the carrier-to-interference-and-noise ratio (CINR) when there is at
least one interference signal present.  This paper assumes a single interference
signal and source at each epoch.  Let CINR$_{ij}$ denote the true CINR for
$i \in \mathcal{I}$ and $j \in \mathcal{J}$, expressed as
\begin{align}
  \label{eq:CINRij}
  \text{CINR}_{ij} &= P^{ij}_\text{R} - 10\,\text{log}_{10} \left( \tilde{N}^i_0 + \tilde{I}^i_0 \right) \quad\quad\text{(dB-Hz)}
\end{align}
where $\tilde{N}^i_0$ is the linear unit equivalent of the noise power density
$N^i_0$, satisfying $N^i_0 = 10\,\text{log}_{10} ( \tilde{N}^i_0 )$, and
$\tilde{I}^i_0$ is the linear unit equivalent of the interference power density
$I^i_0$ (dBW/Hz), satisfying $I^i_0 = 10\,\text{log}_{10} ( \tilde{I}^i_0 )$.
Following \citet{humphreysgnsshandbook}, let $I^i(t)$ be the interference
component of the product between the received signal (including the interference
and thermal noise) and the local replica of the desired signal, and let
$S^i_\text{I}(f)$ be its power spectral density (PSD, dBW/Hz).  Then $I^i_0$ may
be defined as
\begin{align}
  I^i_0 = S^i_\text{I}(0)\quad\quad\text{(dBW/Hz)}
\end{align}
If $\tilde{S}^i_{\text{C}}(f)$ is the PSD of the local replica signal's
spreading code, and $\tilde{S}^i_{r_\text{I}}(f)$ is the PSD of the received
interference signal---both in linear units---and $\hat{f}^i_\text{D}$ is the
receiver's estimate of the desired signal's apparent Doppler frequency (Hz),
then $\tilde{S}^i_\text{I}(f)$, the linear unit equivalent of $S^i_\text{I}(f)$,
is formed by the convolution
\begin{align}
  \label{eq:SIiftilde}
  \tilde{S}^i_\text{I}(f) = \tilde{S}^i_{\text{C}}(f) \star \tilde{S}^i_{r_\text{I}}(f + \hat{f}^i_\text{D})
\end{align}
Similar to (\ref{eq:PijR}), let $P^i_\text{I}$ denote the received interference
signal's power at the $i$th station, given by
\begin{align}
  P^i_\text{I}   &= P_\text{I} + G_\text{I}\left(\theta^{i}_\text{I},  \phi^{i}_\text{I}\right) + G_\text{R}^i\left(\theta^{i\text{I}}_\text{R},  \phi^{i\text{I}}_\text{R}\right) + L_{i\text{I}} \quad\quad\text{(dBW)}
\end{align}
where $P_\text{I}$ is the transmitted interference power and $G_\text{I}$ is the
interference source antenna's gain function.  The arguments of $G_\text{I}$ and
$G_\text{R}^i$, and the quantity $L_{i\text{I}}$ are as before except that they
apply to the interference source rather than to the $j$th GNSS signal.  

At the $k$th time epoch, the receiver at the $i$th station reports the
measurement $z_{ij}[k]$, which is either the measured CNR$_{ij}$ under nominal
operating conditions (the null hypothesis $H_0$), or is the measured CINR$_{ij}$
under interference (the alternate hypothesis $H_1$), for all
$j \in \mathcal{J}$.  The reported measurement $z_{ij}[k]$ is modeled under
$H_0$ and $H_1$ as
\begin{align}
  \label{eq:H0meas}
  H_0:&\quad z_{ij}[k] = \text{CNR}_{ij}[k] + w_{ij}[k] \\
  \label{eq:H1meas}
  H_1 :&\quad z_{ij}[k] = \text{CINR}_{ij}[k] + w_{ij}[k]
\end{align}
where $w_{ij} \sim \mathcal{N}(0,\, \sigma_{ij}^{2})$ is zero-mean additive
white Gaussian noise (AWGN) that models measurement error due to thermal noise,
atmospheric effects, multipath, and other minor effects.

\subsection{Detection}
Determining whether transient interference is present or not at each station,
and across a network of stations, is an exercise in detection theory.  The
problem is to identify an interference event that typically persists for 3 to 5
seconds.  Detection techniques from the nascent transient change detection
literature are well suited to this problem
\citep{guepie2012sequential,moustakides2014multiple, egea2018performance}.
During a transient interference event, the measurement $z_{ij}[k]$ transitions
from $H_0$ to $H_1$---due to a sudden injection of $I^i_0$---and then back to
$H_0$ as the event concludes. An optimal transient change detector, such as
proposed in \citet{guepie2012sequential}, is matched to the interference-induced
pattern in $z_{ij}[k]$. But in cases like the present one in which the
interference interval is unknown \emph{a priori}, no optimal solution is
available.  \citet{egea2018performance} recommend a windowed solution, and
derive certain performance bounds.  The current paper approximates the windowed
solution with a simple differencing technique that is more amenable to analysis
and becomes optimal for single-epoch change durations
\citep{moustakides2014multiple}.

Let $T_\text{max}$ be the maximum duration of the transient event, and
$\Delta t$ be the measurement epoch ($\Delta t = 1$ second in the current
context).  Set $l = \lceil T_\text{max}/\Delta t \rceil + 1 \in \mathbb{N}$ as
the detector's index stride.  Then for the $k$th time-epoch, $i$th station, and
$j$th tracked GNSS signal, let the signal-specific detection statistic be given
by
\begin{align} 
  \xi_{ij}[k] = \tfrac{1}{2} \left(z_{ij}[k + l] - 2z_{ij}[k] + z_{ij}[k - l]\right)
\end{align} 
Under $H_0$ and $H_1$, $\xi_{ij}[k]$ takes on distributions
\begin{align}
  H_0 :  & \quad \xi_{ij}[k] \sim \mathcal{N}(0,\, \tfrac{3}{2}\sigma_{ij}^{2}) \\
  H_1 :  & \quad \xi_{ij}[k] \sim \mathcal{N}(\mu_i,\,\tfrac{3}{2}\sigma_{ij}^{2}) 
\end{align}
where the mean $\mu_i$ is
\begin{align}
  \mu_i = 10\,\text{log}_{10} \left( \frac{\tilde{N}^i_0 +
  \tilde{I}^i_0}{\tilde{N}^i_0}\right)  \quad\quad\text{(dB)}
\end{align}
and can be calculated if all relevant parameters are known.  

Let $\mathcal{J}_i[k] \subset \mathcal{J}$ be the set of GNSS signals tracked by
station $i$ at epoch $k$.  Then the station-specific interference detection
statistic $\Lambda_i[k]$ is the across-signal average of $\xi_{ij}[k]$:
\begin{align}
  \label{eq:siteStat}
  \Lambda_i[k] = \frac{1}{|\mathcal{J}_i[k]|}\sum_{j \in \mathcal{J}_i[k]} \xi_{ij}[k]  
\end{align}
The distribution of $\Lambda_i[k]$ under $H_0$ and $H_1$ is
\begin{align}
  H_0:& \quad \Lambda_i[k] \sim \mathcal{N}(0,\, \sigma_{i}^{2}) \\
  H_1:& \quad \Lambda_i[k] \sim \mathcal{N}(\mu_i,\, \sigma_{i}^{2})
\end{align}
The quantity $\mu_i$ amounts to the average drop in observed CNR at the $i$th
station due to the interference event, in dB.  In practice, the parameters that
contribute to $\tilde{I}^i_0$ are unknown \emph{a priori}, leaving $\mu_i$
unknown.  But because $\tilde{I}^i_0 > 0$, it follows that $\mu_i > 0$, which
allows it to be cancelled from each site's detection statistic, making a
test based on (\ref{eq:siteStat}) uniformly most powerful \citep{vtrees2001dem}.
The value $\sigma_{i}^{2}$ can be estimated at the $i$th station using
historical data, and from this estimate a detection threshold $\nu_i$ can be
calculated for a constant false alarm rate (CFAR).  The hypothesis test for the
$i$th station then becomes
\begin{align}
  \label{eq:hypTest}
  \Lambda_i[k] \mathop{\gtrless}_{H_0}^{H_1} \nu_i
\end{align}

\ifpreprint
	\begin{figure}[b]
	\centering
	\begin{minipage}[b]{0.48\textwidth}
		\centering
		\includegraphics[width=\linewidth]{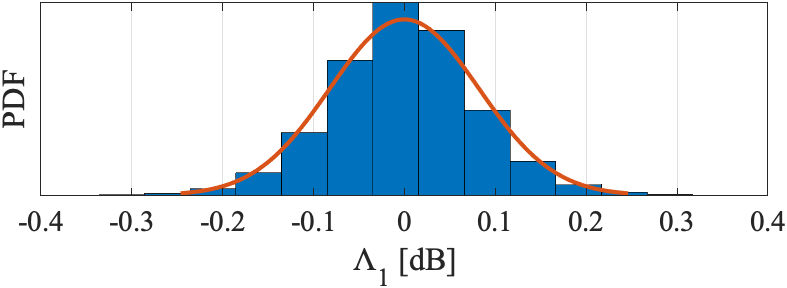}
	\end{minipage}
	\hfill
	\begin{minipage}[b]{0.48\textwidth}
		\centering
		\includegraphics[width=\linewidth]{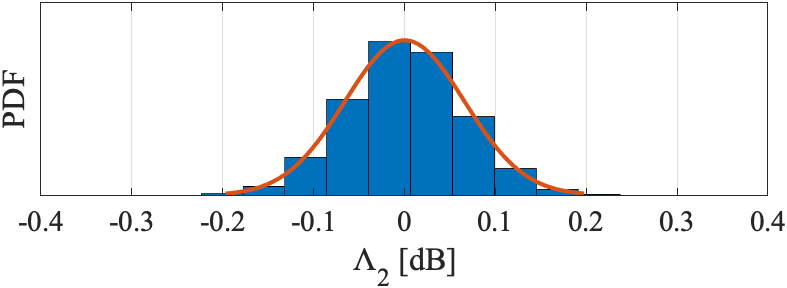}
	\end{minipage}
	\caption{The distributions of $\Lambda_i$ under $H_0$ (no interference
		present) for stations METG ($i = 1$) and MATE ($i = 2$).}
	\label{fig:nullHypothesis}
\end{figure}
\else
	\begin{figure}[t]
	\centering
	\begin{minipage}[b]{0.48\textwidth}
		\centering
		\includegraphics[width=\linewidth]{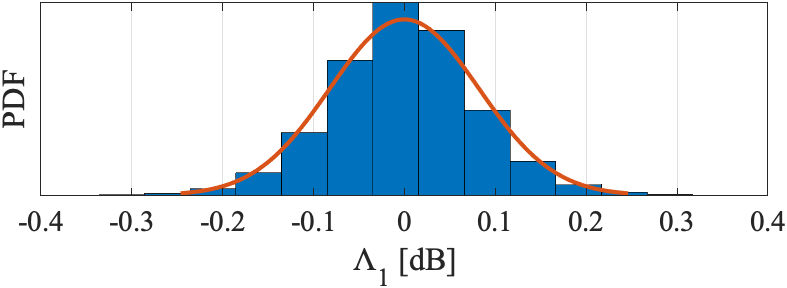}
	\end{minipage}
	\hfill
	\begin{minipage}[b]{0.48\textwidth}
		\centering
		\includegraphics[width=\linewidth]{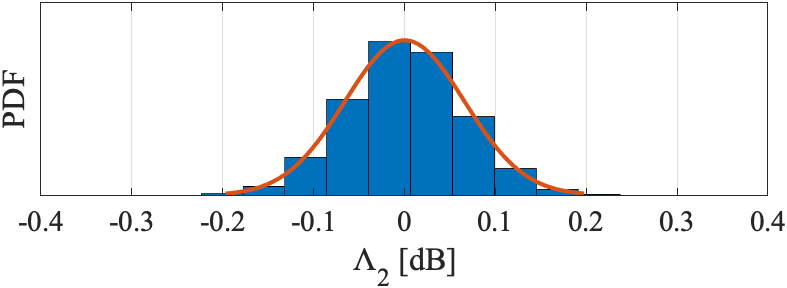}
	\end{minipage}
	\caption{The distributions of $\Lambda_i$ under $H_0$ (no interference
		present) for stations METG ($i = 1$) and MATE ($i = 2$).}
	\label{fig:nullHypothesis}
\end{figure}
\fi

Null-hypothesis ($H_0$) detection statistic distributions from a day's worth of
data for two stations, METG and MATE, are shown in
Fig.~\ref{fig:nullHypothesis}. Both appear to be zero mean and approximately
Gaussian distributed.  The modest difference in $\sigma_i^2$ can be attributed
to local noise, different receivers and antennas, and quantization of the
reported CNR.  

The remainder of this section focuses on day 160 of year
2021. Fig.~\ref{fig:CN0} shows the time history of $\{z_{ij}\}$ and
$\{\Lambda_i\}$ over a 15-minute interval for three stations with $l = 3$,
amounting to a 3-second stride at the 1-Hz measurement rate. The dashed red
lines in the right plots indicate the detection thresholds corresponding to a
$10^{-4}$ probability of false alarm.  Around the 700-second mark, interference
is detected simultaneously at all three stations.  While individual interference
detections are not uncommon because reference stations occasionally experience
local interference, time alignment of an interference event across
geographically distant sites in Finland, Italy, and Greenland is noteworthy.
Moreover, the interference exhibits the same characteristic signature at each
location: a CNR drop lasting approximately 3 seconds. This suggests that all
three stations were affected by a common interference source.
\ifpreprint
	\begin{figure}[t]
  \centering
  \begin{minipage}[b]{0.49\textwidth}
    \centering
    \includegraphics[width=\linewidth]{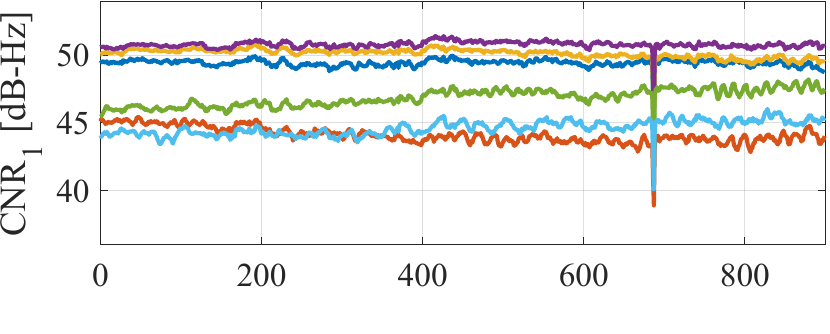}
    \includegraphics[width=\linewidth]{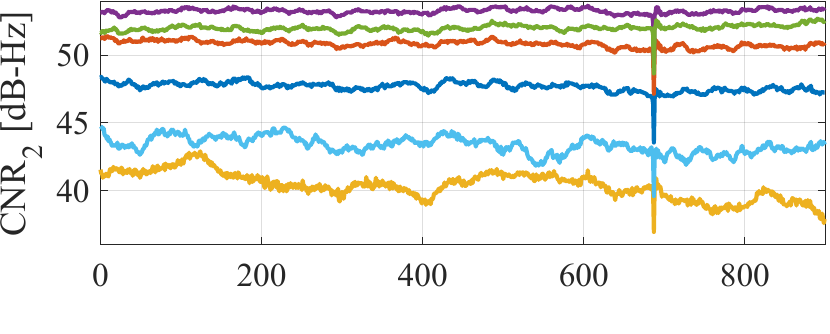}
    \includegraphics[width=\linewidth]{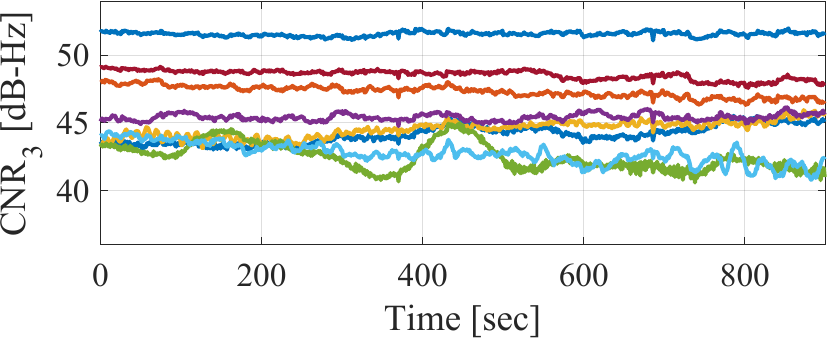}
  \end{minipage}
  \begin{minipage}[b]{0.2\textwidth}
    
  \end{minipage}
  \hfill
  \begin{minipage}[b]{0.49\textwidth}
    \centering
    \includegraphics[width=\linewidth]{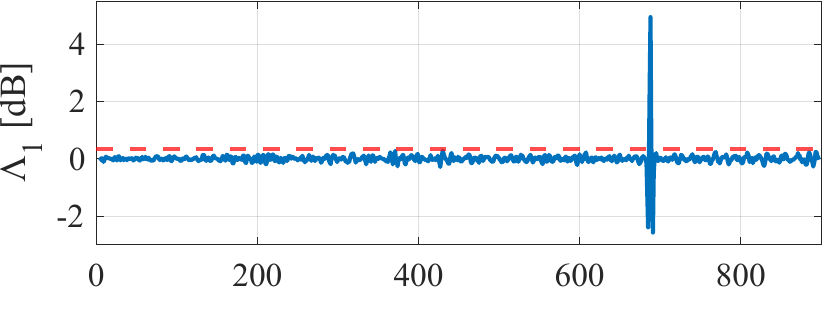}
    \includegraphics[width=\linewidth]{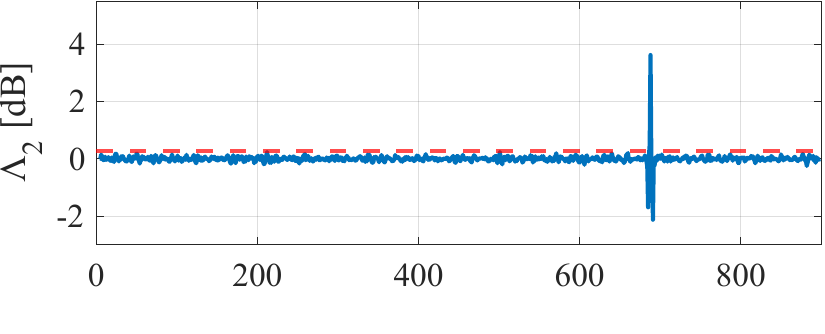}
    \includegraphics[width=\linewidth]{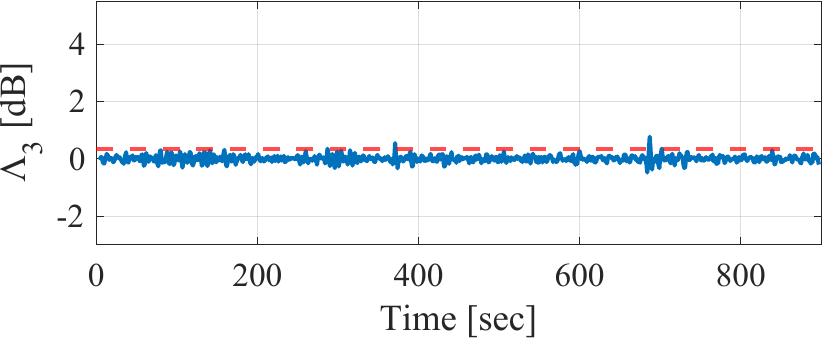}
  \end{minipage}
  \caption{Reported CNR (left) and the detection statistic $\Lambda_i$
    (right) for IGS stations METG ($i = 1$, Finland), MATE ($i = 2$,
    Italy), and THU2 ($i = 3$, Greenland) over a 15-minute interval on day
    160 of year 2021.  The dashed red line is the detection threshold with
    a $10^{-4}$ probability of false alarm.}
  \label{fig:CN0}
\end{figure}
\else
	\begin{figure}[t]
	\centering
	\begin{minipage}[b]{0.48\textwidth}
		\centering
		\includegraphics[width=\linewidth]{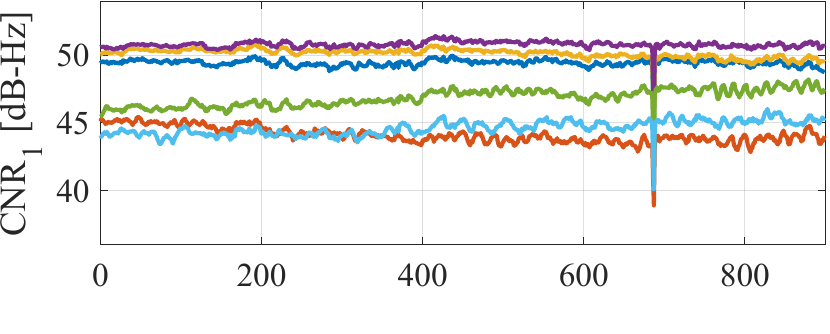}
		\includegraphics[width=\linewidth]{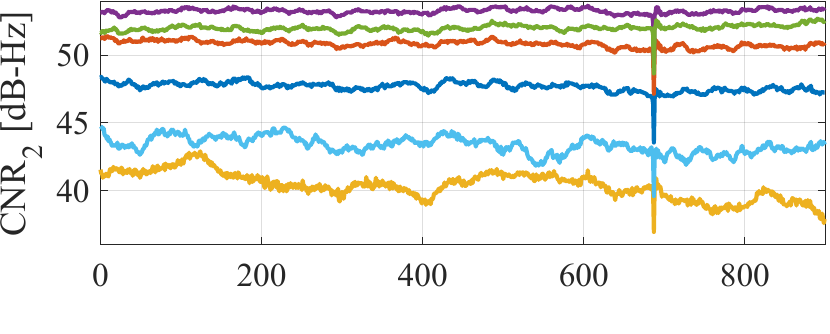}
		\includegraphics[width=\linewidth]{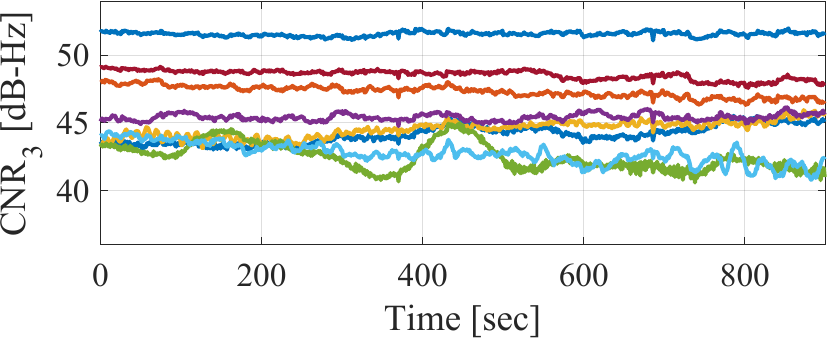}
	\end{minipage}
	\begin{minipage}[b]{0.2\textwidth}
		
	\end{minipage}
	\hfill
	\begin{minipage}[b]{0.48\textwidth}
		\centering
		\includegraphics[width=\linewidth]{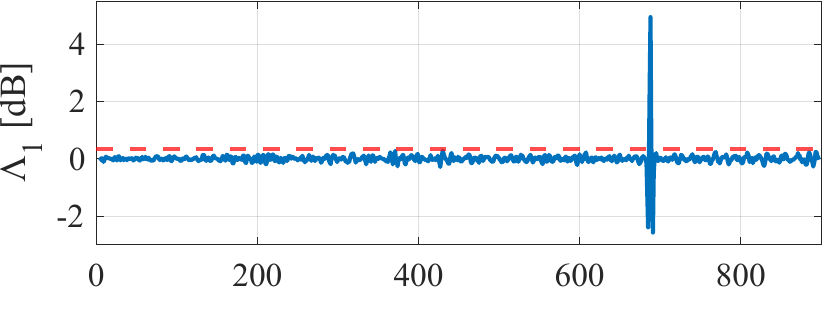}
		\includegraphics[width=\linewidth]{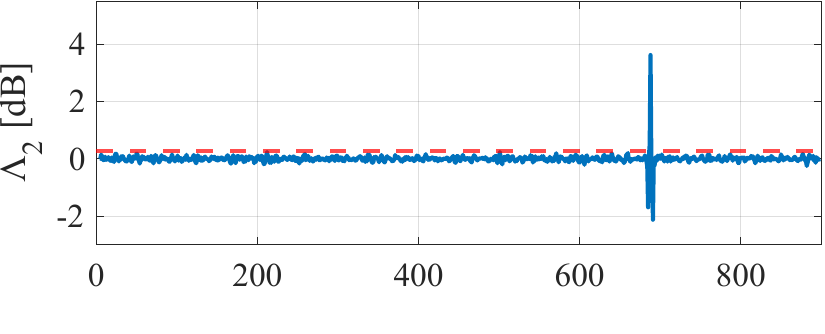}
		\includegraphics[width=\linewidth]{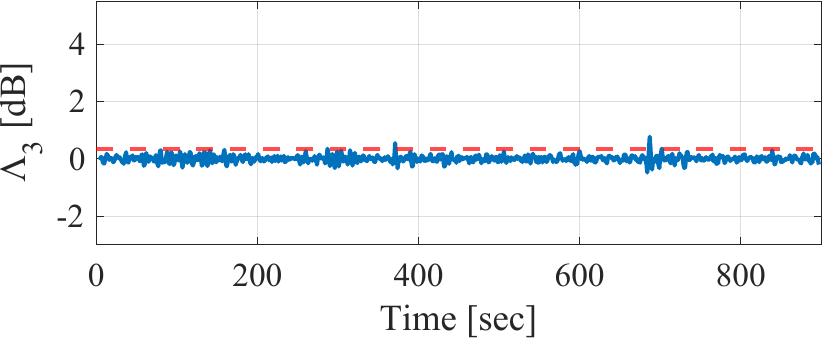}
	\end{minipage}
	\caption{Reported CNR (left) and the detection statistic $\Lambda_i$
		(right) for IGS stations METG ($i = 1$, Finland), MATE ($i = 2$,
		Italy), and THU2 ($i = 3$, Greenland) over a 15-minute interval on day
		160 of year 2021.  The dashed red line is the detection threshold with
		a $10^{-4}$ probability of false alarm.}
	\label{fig:CN0}
\end{figure}
\fi
\begin{figure}[t]
  \centering
  \begin{minipage}[b]{0.48\textwidth}
    \centering
    \includegraphics[width=\linewidth]{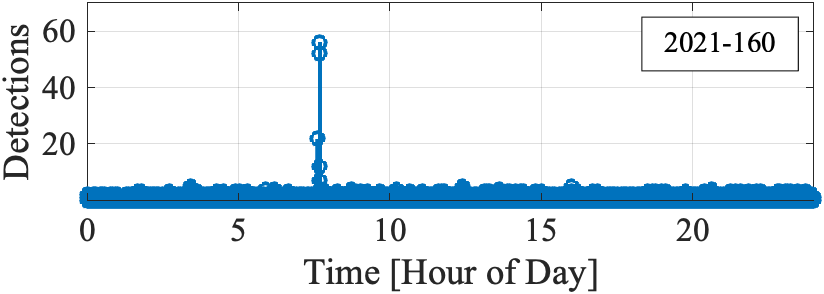}
  \end{minipage}
  \hfill
  \begin{minipage}[b]{0.48\textwidth}
    \centering
    \includegraphics[width=\linewidth]{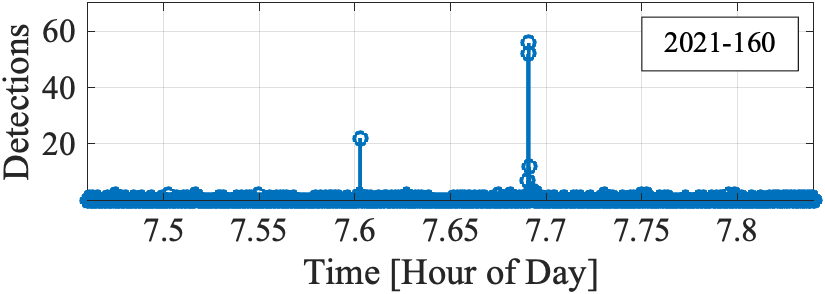}
  \end{minipage}
  \caption{The number of stations that detected interference during day
    160 of year 2021.  The expanded view on the right shows a lower-power
    event detected by 21 stations, followed by a higher-power event
    detected by 58 stations.}
  \label{fig:detection}
\end{figure}

The detection test was conducted continuously across stations spanning Europe,
Greenland, and Canada on day 160 of 2021. Fig.~\ref{fig:detection} shows the
number of stations registering detections during this period with a 10$^{-4}$
probability of false alarm. A lower-power interference event was detected by 21
stations, followed by a more powerful event detected by 58 stations. The
stronger event corresponds to the 700-second mark shown in
Fig.~\ref{fig:CN0}. The recorded CNR observables at each station align with the
patterns depicted in Fig.~\ref{fig:CN0}, providing further evidence that these
stations were affected by the same interference source. A heat map showing the
spatial distribution of the CNR reduction is presented in
Fig.~\ref{fig:2021160drop}. During the more powerful event, the tracked CNR of
GPS L1 C/A signals exhibited drops as large as 6 dB, while the weaker event
produced drops up to 1.5 dB.
\ifpreprint
	\begin{figure}[t]
	  \centering
	  \begin{minipage}[b]{0.44\textwidth}
	    \centering
	    \includegraphics[width=\linewidth]{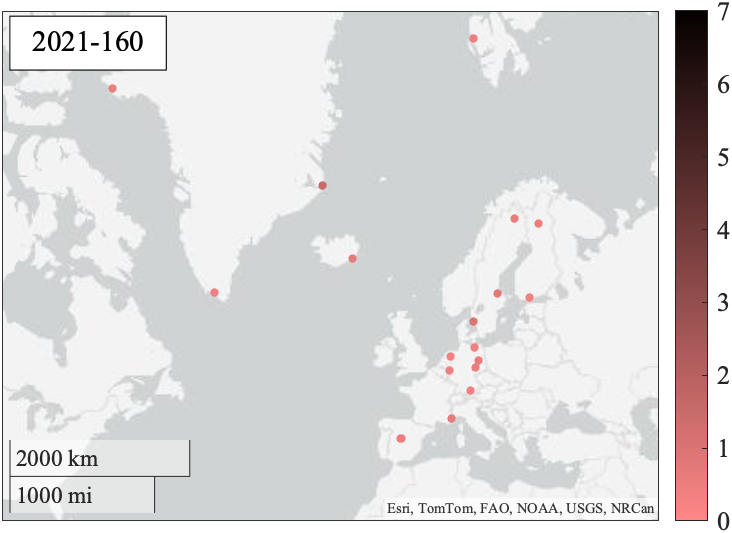}
	  \end{minipage}
	  \hspace{5mm}
	  \begin{minipage}[b]{0.44\textwidth}
	    \centering
	    \includegraphics[width=\linewidth]{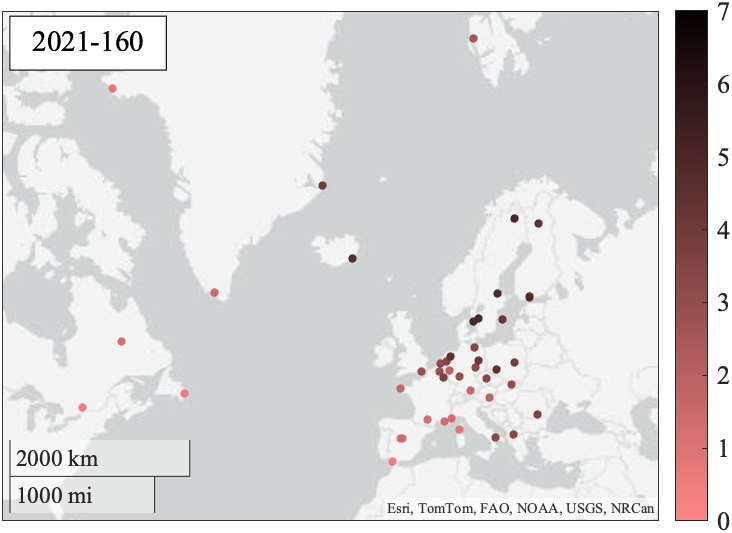}
	  \end{minipage}
	  \caption{Heat map of the test statistic (in dB) at triggering stations
	    for the lower-power event (left) and the higher-power event (right).
	    The drop in GPS L1 C/A CNR during the more powerful interference event
	    was as large as 6 dB, centered near the Baltic region.}
	  \label{fig:2021160drop}
	\end{figure}
\else
	\begin{figure}[t]
	\centering
	\begin{minipage}[b]{0.49\textwidth}
		\centering
		\includegraphics[width=\linewidth]{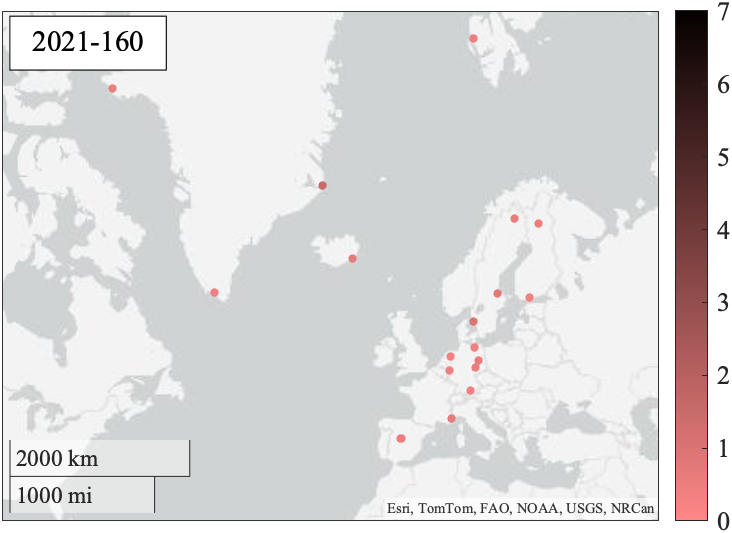}
	\end{minipage}
	\hfill
	\begin{minipage}[b]{0.49\textwidth}
		\centering
		\includegraphics[width=\linewidth]{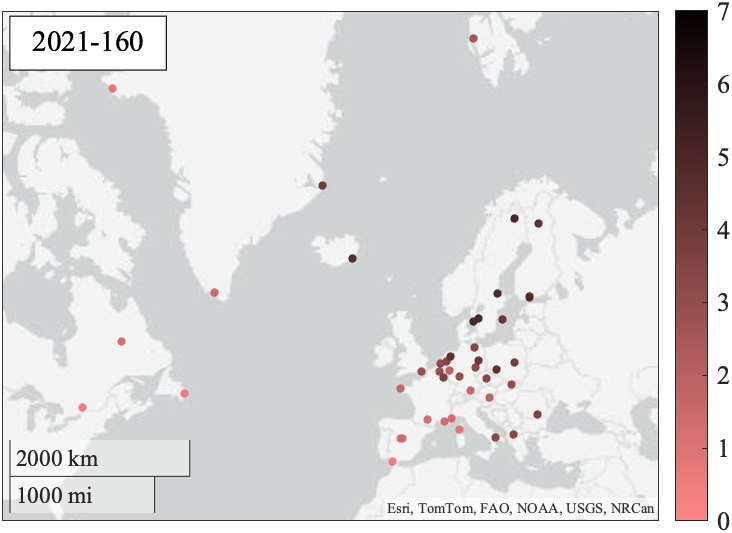}
	\end{minipage}
	\caption{Heat map of the test statistic (in dB) at triggering stations
		for the lower-power event (left) and the higher-power event (right).
		The drop in GPS L1 C/A CNR during the more powerful interference event
		was as large as 6 dB, centered near the Baltic region.}
	\label{fig:2021160drop}
\end{figure}
\fi

\ifpreprint
\newpage
\fi
\section{Interference Properties}
This section details the transient space-based interference's temporal, spatial,
and spectral properties over the seven-year period from January 2019 to April
2026.  For comparison, it also presents data from a naturally occurring solar
radio burst.

\subsection{Temporal Patterns}
\begin{figure}[t]
  \centering
  \begin{minipage}[b]{0.45\textwidth}
    \centering
    \includegraphics[width=\linewidth]{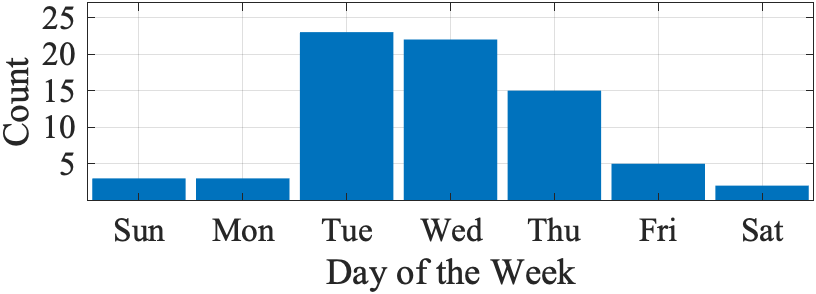}
  \end{minipage}
  \hfill
  \begin{minipage}[b]{0.45\textwidth}
    \centering
    \includegraphics[width=\linewidth]{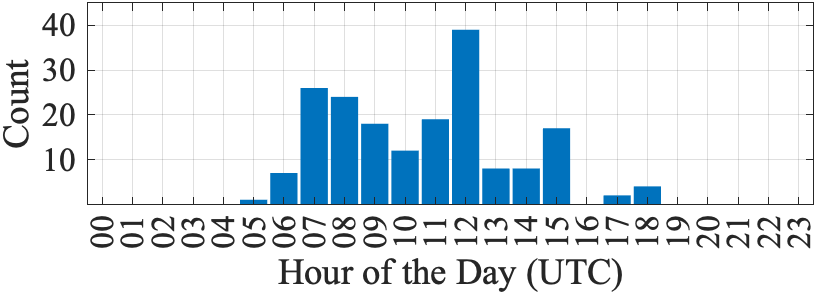}
  \end{minipage}
  \caption{Distribution of the day of the week and hour of day (with
    respect to UTC) during which interference events with at least one
    station suffering a drop of 5 dB or greater occurred.  Clearly, the
    high-power interference events typically occur during business days
    and business hours.}
  \label{fig:human}
\end{figure}

High-rate (1-Hz) data from 165 reference stations collected over the seven-year
period were retrieved and analyzed. The detection hypothesis test described
earlier was applied to every station on an epoch-by-epoch basis from January 1,
2019 to May 4, 2026. If interference was detected simultaneously at several
stations, then an event detection was declared. A total of 75 days were
identified with at least one wide-area transient GNSS interference event on the
GPS L1 frequency during which at least one station experienced a CNR drop of 5
dB or greater. The annual number of such occurrences is summarized in
Table~\ref{tbl:occurances}. The earliest detection of a significant transient
wide-area interference event within this period occurred in October 2019.
Fig.~\ref{fig:human} presents a histogram showing the distribution of high-power
interference events by day of the week and hour of the week.  Notably, these
predominantly occurred during business days and business hours (UTC time), which
suggests human involvement: a purely random phenomenon would tend to exhibit a
temporally uniform distribution.

\ifpreprint
\begin{table}[b]
	\caption{The number of days per year that saw at least one wide-area transient
		GNSS interference event on the GPS L1 frequency with at least one station
		suffering a drop of 5 dB or greater.}
	\centering
	\normalsize
	\begin{tabular}{ c | c c c c c c c c | c } 
		\toprule
		& 2019 & 2020 & 2021 & 2022 & 2023 & 2024 & 2025 & 2026 & Total \\ 
		\hline
		Count & 8 & 17 & 5 & 7 & 15 & 9 & 12 & 2 & 75 \\ 
		\bottomrule
	\end{tabular}
	\label{tbl:occurances}
\end{table}
\fi

There were an 47 additional days with weaker widespread transient interference
events.  On these days, the largest CNR drop did not exceed 5 dB. If these days
are included in Fig.~\ref{fig:human}, the distributions shown do not change
appreciably. The weaker interference events are left out of this paper's
analysis to avoid false alarms at the periphery of the affected area.

\unless\ifpreprint
\begin{table}[t]
  \caption{The number of days per year that saw at least one wide-area transient
    GNSS interference event on the GPS L1 frequency with at least one station
    suffering a drop of 5 dB or greater.}
  \centering
  \normalsize
  \begin{tabular}{ c | c c c c c c c c | c } 
    \toprule
    & 2019 & 2020 & 2021 & 2022 & 2023 & 2024 & 2025 & 2026 & Total \\ 
    \hline
    Count & 8 & 17 & 5 & 7 & 15 & 9 & 12 & 2 & 75 \\ 
    \bottomrule
  \end{tabular}
  \label{tbl:occurances}
\end{table}
\fi

\begin{figure}[t]
  \centering
  \begin{minipage}[b]{0.48\textwidth}
    \centering
    \includegraphics[width=\linewidth]{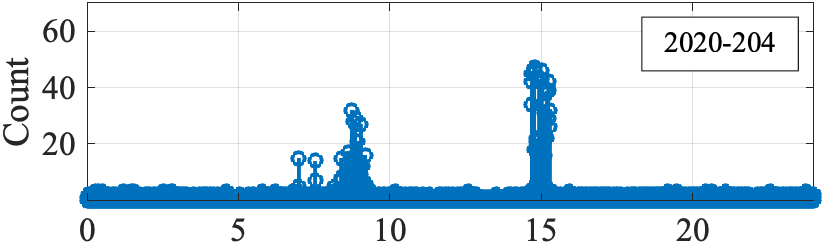}
    
    \vspace{2mm}
    
    \includegraphics[width=\linewidth]{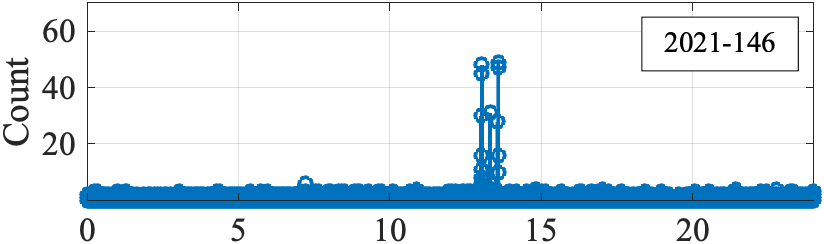}
    
    \vspace{2mm}
    
    \includegraphics[width=\linewidth]{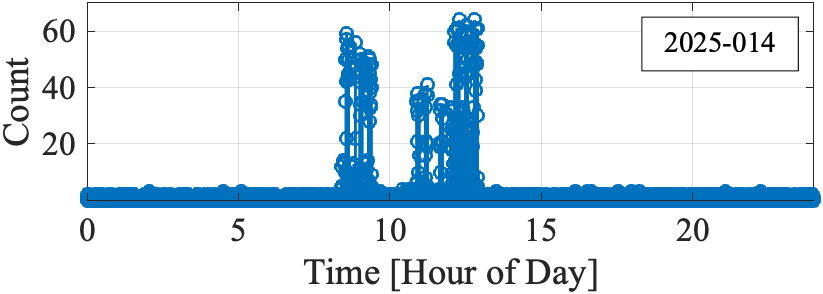}
  \end{minipage}
  \hfill
  \begin{minipage}[b]{0.48\textwidth}
    \centering
    \includegraphics[width=\linewidth]{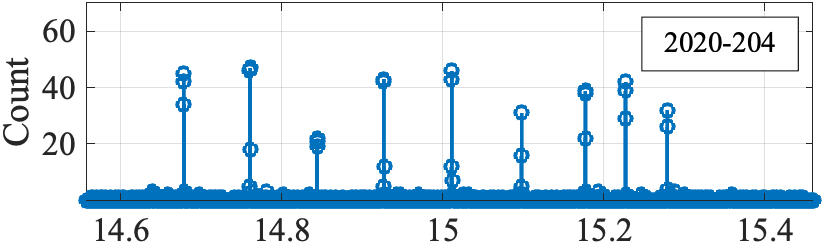}
    
    \vspace{2mm}
    
    \includegraphics[width=\linewidth]{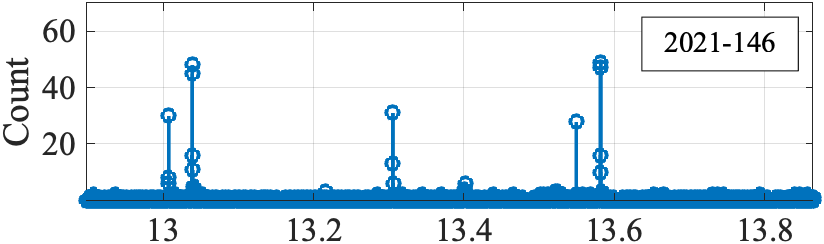}
    
    \vspace{2mm}
    
    \includegraphics[width=\linewidth]{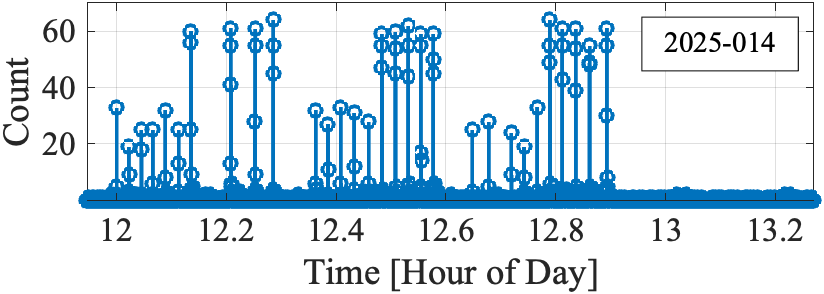}
  \end{minipage}
  \caption{The number of stations that detected interference during three
    different days.  Interference events can occur several times a day. The left
    column shows the entire day, while the right column is an expanded
    sub-interval. }
  \label{fig:detections}
\end{figure}

Fig.~\ref{fig:detections} presents the number of daily detections across three
additional example days. Unlike the case in Fig.~\ref{fig:detection}, which
shows a single high-power interference event, some days experienced multiple
high-power events. Notably, day 146 of 2021 exhibited an interference signature
similar to that of day 160 of the same year, with a low-power burst followed by
a high-power burst. This happened twice on day 146, with the strong interference
events being separated by approximately 32.6 minutes.  The time delay between
the low-power and high-power burst was 317 seconds on day 160, and 115 seconds
for both events on day 146.  This signature was observed on several other days.
Day 204 of 2020 and day 014 of year 2025 recorded numerous high-power
interference bursts. The timing of the bursts within a single day is typically
periodic, with large bursts often spaced by an integer multiple of 150 seconds.
The daily temporal patterns shown here of the high-power interference bursts are
broadly representative of the phenomenon across all 75 days.

\subsection{Spatial Patterns}
\begin{figure}[t]
  \centering
  \begin{minipage}[b]{0.33\textwidth}
    \centering
    \includegraphics[width=\linewidth]{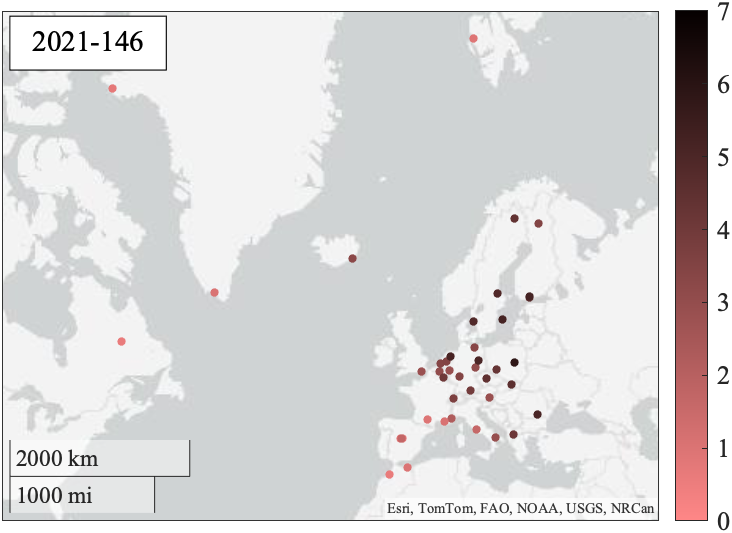}
  \end{minipage}
  \begin{minipage}[b]{0.33\textwidth}
    \centering
    \includegraphics[width=\linewidth]{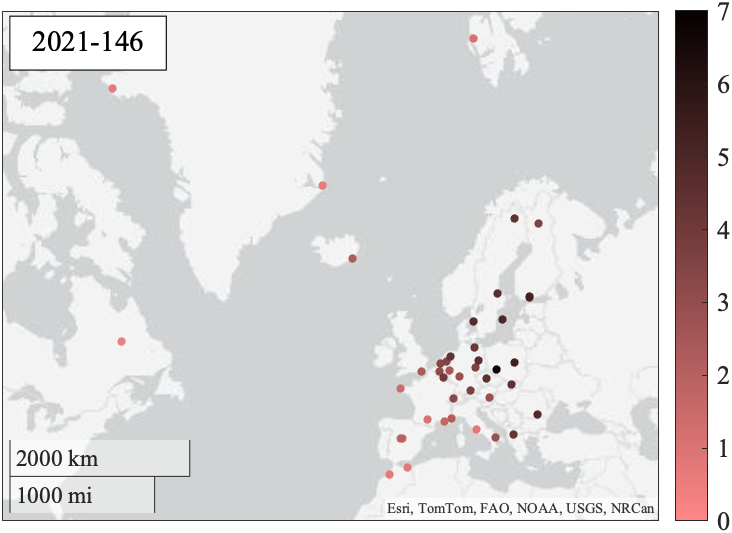}
  \end{minipage}
  \begin{minipage}[b]{0.33\textwidth}
    \centering
    \includegraphics[width=\linewidth]{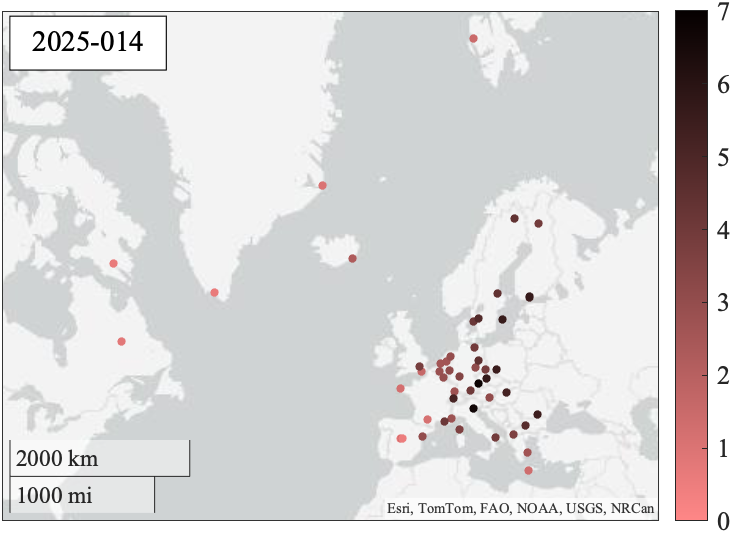}
  \end{minipage}
  \caption{Detection statistic heat maps, in dB, for day 146 of year 2021 and
    day 014 of year 2025.  GNSS receivers in the Baltic region are impacted
    most, which is generally representative of nearly all of the interference
    events. }
  \label{fig:heatmap2}
\end{figure}

Across all 75 days on which high-power events occurred, GNSS receivers in Europe
were the most affected, with the Baltic region consistently experiencing the
largest CNR drops.  Fig.~\ref{fig:heatmap2} shows example heat maps from day 146
of 2021 and day 014 of 2025. On day 146, two high-power interference events
occurred, both producing nearly identical spatial patterns. Day 014 of 2025 also
saw multiple high-power bursts, all with consistent patterns matching the
example shown. The largest CNR drop across all events was 10 dB recorded at the
LAMA station in Poland in 2025. Notably, during wide-area interference events in
Europe, no similar disruptions were detected elsewhere in the world.

Although the overwhelming majority of interference events saw receivers in the
Baltic region impacted most, day 204 of 2020 exhibited a distinct interference
pattern compared to other days.  On this day, there was progressive movement in
the geographic center of interference over multiple events, starting in the
Baltic Sea and then moving into Germany and on to the Norwegian Sea, all over a
20-minute interval \citep{clements2025spaceintgnss}.  This deviation may be
attributed to satellite motion, to a change in the interference source's beam
pointing vector, or to multiple active satellite sources.

\subsection{Spectral Properties}
Spectral data were collected using a u-blox F9P GNSS receiver connected to a
Trimble GNSS antenna installed in Gdynia,~Poland, operating as an experimental
radio frequency interference (RFI) monitoring station.  Direct estimation of the
absolute received interference power at the antenna reference point is not
possible because of the unknown loss between the antenna and the receiver front
end.  Accordingly, the analysis here focuses on relative comparisons of the PSD
across interference events and against interference-free baseline
conditions. The u-blox receiver provides uncalibrated 1-Hz spectral observations
through the ``SPAN” message.  Although SPAN output measurements are
dimensionless, they are nonetheless suitable for comparative analysis. In
previous work, a methodology was developed to transform these uncalibrated SPAN
observations into a spectrally adjusted power density metric
\citep{kriezis2025gnss}.

Fig.~\ref{fig:spectrum} presents the raw SPAN spectral output under nominal
(RFI-free) conditions (black line) and during 48 strong interference events from
2024 to 2025.  The interference spectrum exhibits a consistent spectral shape
across all recorded events on days for which data are available. The
interference peak is centered at 1577.5 MHz, about 2 MHz above the GPS L1 center
frequency of 1575.42 MHz, and has an approximate bandwidth of 5 MHz. 

\unless\ifpreprint
\begin{figure}[t]
  \centering
  \includegraphics[width=.87\linewidth]{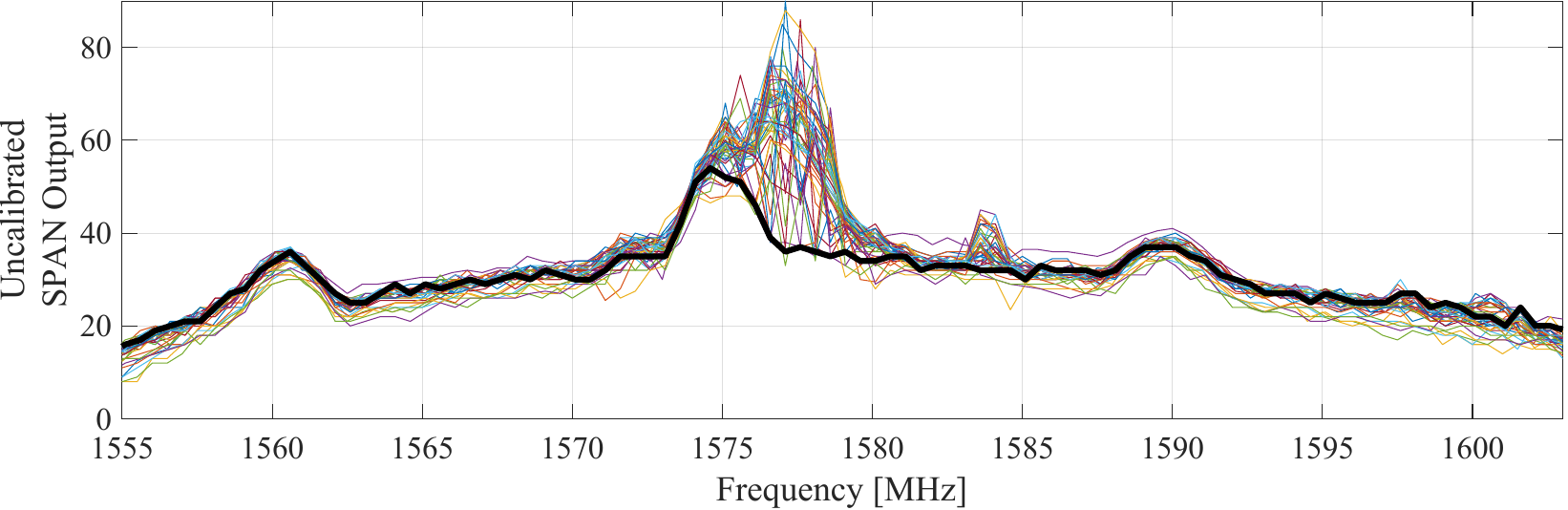}
  \caption{Uncalibrated average PSD near the GPS L1 band during nominal
    operation (thick black line) and 1-Hz PSD estimates during 48 transient
    interference events, as recorded in Gdynia, Poland.}
  \label{fig:spectrum}
\end{figure}
\fi
In addition to tracked GPS L1 C/A signals, tracked Galileo E1 and BeiDou
B1C/B1A signals also exhibited a concurrent drop in CNR during interference
events.  The magnitude of the decrease at each station closely matched the drop
of the GPS L1 C/A signals, which might be expected because they share the same
center frequency.  The drop in CNR is not identical because the GPS, Galileo,
and BeiDou spreading codes each have a different PSD $\tilde{S}^i_{\text{C}}(f)$
which, as shown in (\ref{eq:SIiftilde}), leads to a different interference
component PSD $\tilde{S}^i_\text{I}(f)$ and ultimately to a different
interference power density $I_0^i$.

During strong 1577.5-MHz-centered interference events, a small but noticeable
drop in CNR is also observed on tracked BeiDou B1I signals, which are centered
at 1561.098 MHz.  This is a puzzling because the interference spectrum shown in
Fig.~\ref{fig:spectrum} has no spectral overlap with the BeiDou B1I signal, from
which it follows that $I_0^i = S^i_\text{I}(0) = 0$.  Apparently, the method by
which receivers in the IGS network estimate the noise floor $N_0^i$ that they
use to report CNR values is somewhat sensitive to interference near L1.

Occasionally, an initial CNR drop and recovery in tracked GPS L1 C/A signals was
followed by an equal-magnitude CNR drop and recovery in tracked BeiDou B1I
signals. This pattern was observed on 15 of the 75 days, the first in June
2020. This indicates that the interference source can generate signals at 1577.5
MHz and near 1561.098 MHz.  Fig.~\ref{fig:spectrum} only shows the interference
centered at 1577.5 MHz, but Fig.~\ref{fig:PSD}, which shows PSDs derived from
raw wideband samples captured in February 2026 by a receiver in the Netherlands
clearly shows both interference bands, with the lower band centered at 1558.5
MHz.  Interestingly, these two interference bands never appear to be active at
the same time.

No wide-area transient interference has yet been observed near the GPS L2 or L5
bands.

\unless \ifpreprint
\begin{figure}[t]
  \centering
  \includegraphics[width=.85\linewidth]{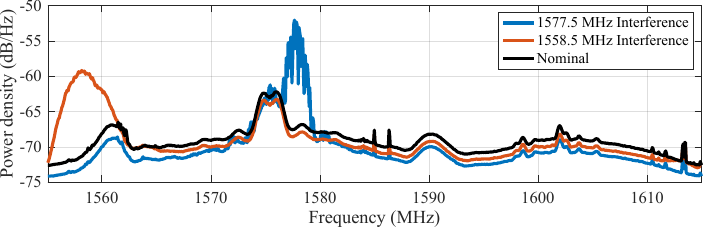}
  \caption{Power spectrum derived from raw wideband samples captured in
    Amsterdam, Netherlands, during an interference event on February 11, 2026.
    An initial burst of interference at 1577.5 MHz was followed by a burst at
    1558.5 MHz.}
  \label{fig:PSD}
\end{figure}
\fi

\ifpreprint
\begin{figure}[t]
	\centering
	\includegraphics[width=.98\linewidth]{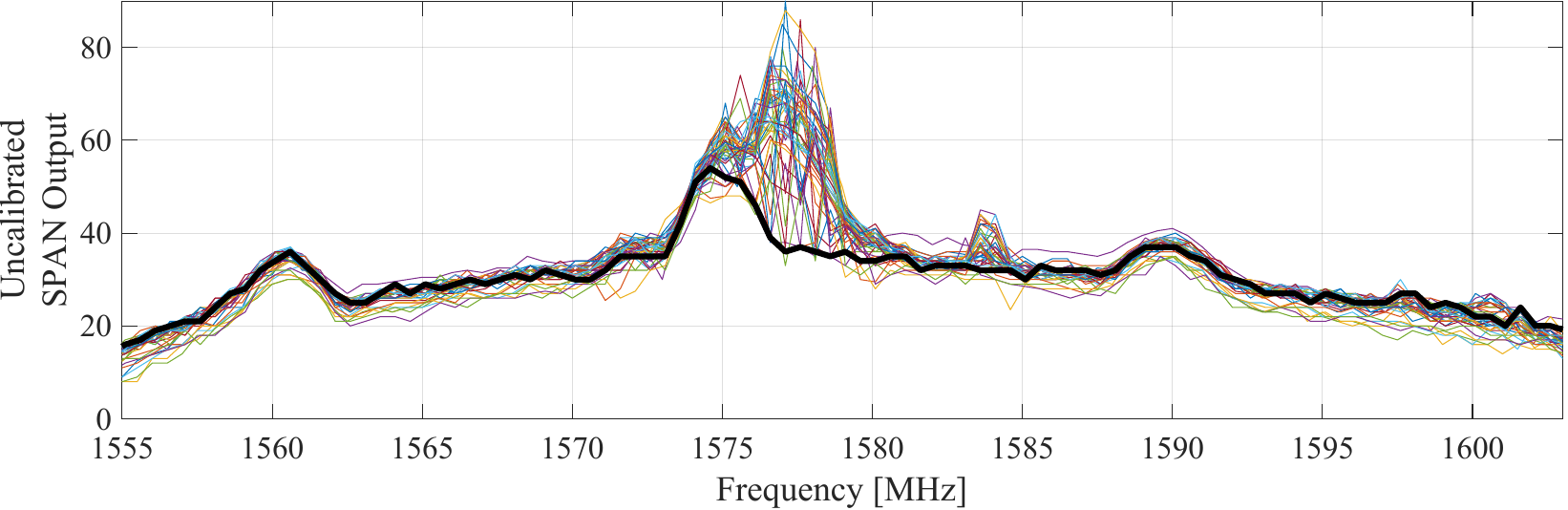}
	\caption{Uncalibrated average PSD near the GPS L1 band during nominal
		operation (thick black line) and 1-Hz PSD estimates during 48 transient
		interference events, as recorded in Gdynia, Poland.}
	\label{fig:spectrum}
\end{figure}

\begin{figure}[t]
	\centering
	\includegraphics[width=.95\linewidth]{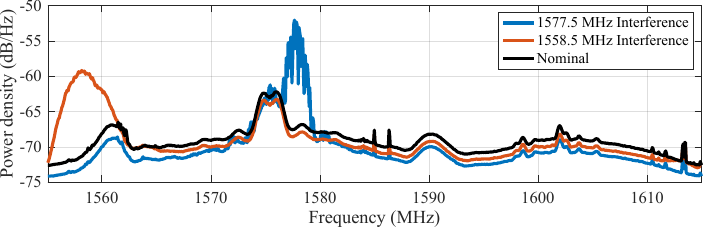}
	\caption{Power spectrum derived from raw wideband samples captured in
		Amsterdam, Netherlands, during an interference event on February 11, 2026.
		An initial burst of interference at 1577.5 MHz was followed by a burst at
		1558.5 MHz.}
	\label{fig:PSD}
\end{figure}
\fi

\subsection{Comparison with Solar Radio Burst}
It is worth noting that naturally occurring phenomena can also cause significant
CNR reduction over a large geographic area. \citet{cerruti2006srb,
  cerruti2008effect} examined the impact of intense solar radio bursts on
tracked GNSS signals using observational data from the IGS network during solar
events in the mid 2000s. It was demonstrated that solar radio bursts can cause
CNR degradation by as much as 25 dB across the sunlit side of Earth. But solar
radio bursts are qualitatively different from the transient interference studied
in this paper: They are typically broadband, evolve more slowly than the rapid
onset observed in Fig.~\ref{fig:detection}, and the associated CNR reduction
typically persists for a longer duration.

\begin{figure}[t]
  \centering
  \begin{minipage}[b]{0.42\textwidth}
    \centering
    \includegraphics[width=\linewidth]{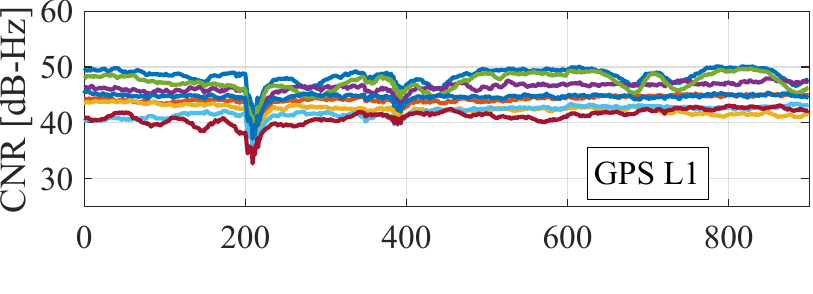}
    \includegraphics[width=\linewidth]{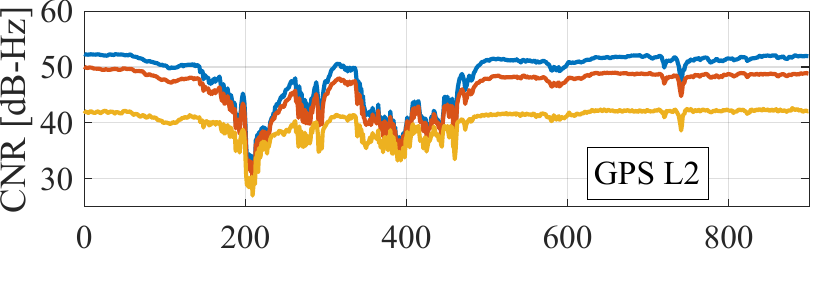}
    \includegraphics[width=\linewidth]{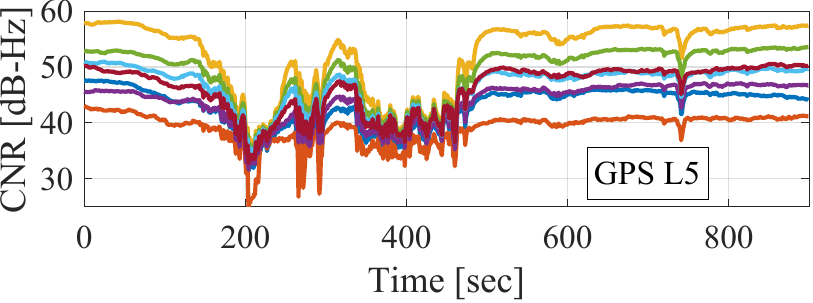}
  \end{minipage}
  \hspace{5mm}
  \begin{minipage}[b]{0.495\textwidth}
    \centering
    \includegraphics[width=\linewidth]{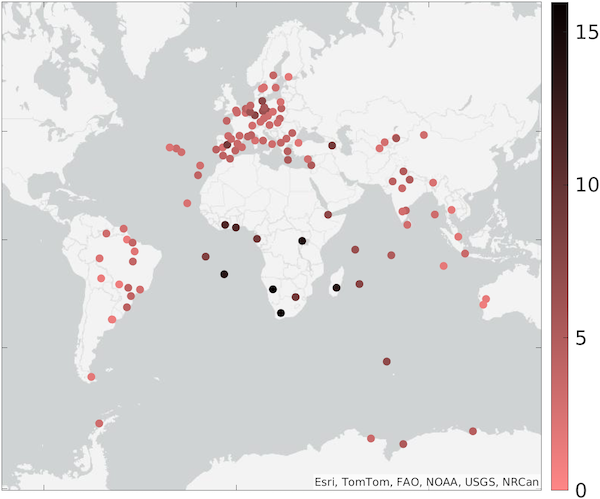}
    
    \vspace{8mm}
    
  \end{minipage}
  \caption{Left: CNR time history of tracked GPS L1, L2, and L5 signals produced
    by IGS station SUTM in South Africa during the solar radio burst on November
    11, 2025.  Right: Heat map of maximum CNR degradation at GPS L2 during the
    solar radio burst.}
  \label{fig:solarEvent}
\end{figure}

On November 11, 2025, a strong solar flare occurred with a magnitude of X5.1,
causing a geomagnetic storm on Earth reaching category G4 (``severe")
\citep{ESA2025solarStorm}.  IGS stations positioned on the sunlit side of Earth
experienced drops in CNR on all GNSS frequencies. Fig.~\ref{fig:solarEvent}
shows the CNR time history of tracked GPS L1, L2, and L5 signals from the SUTM
reference station in South Africa.  Signals on L2 and L5 were degraded as much
as 17 dB for hundreds of seconds.  The CNR of tracked signals at other IGS
stations were degraded in a similar manner. Fig.~\ref{fig:solarEvent} shows the
magnitude of the CNR drop on the tracked L2 signals across the globe.  Clearly,
the effects of solar radio bursts manifest differently in the IGS data compared
to the transient phenomenon studied here.

\section{Elevation-Mask-Based Interference Source Identification}
\label{sec:basicsatid}
A simple technique for winnowing the list of possible candidate satellites
causing interference is to determine which satellites were overhead the affected
region during a given interference event. Two-Line Elements (TLEs) for publicly
tracked orbiting objects can be obtained from space-track.org, a database
maintained by the United States Space Force to support spaceflight safety. TLEs
can be used to compute the approximate position of an object at any time
reasonably close to the TLE epoch. While the accuracy of these estimates is
roughly 1 km at epoch and degrades with the age of the TLE
\citep{komodromos2025networkplans, morgan2025mockplans}, their precision is more
than sufficient for a preliminary candidacy test.

Let $\mathcal{S}$ be the set of all space objects with public TLEs, and
$\mathcal{I}_\text{d} \subseteq \mathcal{I}$ be the set of all reference
stations that detected a given interference event.  The elevation angle of
satellite $s \in \mathcal{S}$ as seen from the GNSS antenna at station
$i \in \mathcal{I}_\text{d}$ is denoted $\alpha_{is}$.  This can be calculated
assuming an ellipsoidal model for Earth such as WGS84.  Assuming an elevation
mask $\alpha_0$, satellite $s \in \mathcal{S}$ is considered a valid candidate
if $\alpha_{is} \geq \alpha_0$ for all $i \in \mathcal{I}_\text{d}$.

Under this framework, one can readily determine the minimum altitude at apogee
for candidate satellites.  Let $\vb{r} \in \mathbb{R}^3$ be an arbitrary
position in Earth-centered, Earth-fixed (ECEF) coordinates, and
$\alpha_i(\vb{r})$ be its elevation angle as seen from station
$i \in \mathcal{I}_\text{d}$.  The minimum-radius position $\vb{r}^*$ satisfying all
elevation mask constraints may be found by solving
\begin{align}
  \label{eq:nlrMin}
  \vb{r}^* = \argmin_{\vb{r} \in \mathbb{R}^3} \|\vb{r}\| \quad \text{such that}
  \quad \alpha_i(\vb{r}) \geq \alpha_0 \quad \text{for all} \quad i \in \mathcal{I}_\text{d}
\end{align}
A solution, which is unique for $\alpha_0 \geq 0$, can be obtained
straightforwardly using a numerical optimizer. The minimum satellite altitude at
apogee is then given by $\zeta^* = \| \vb{r}^* \| - r_\text{E}$, where
$r_\text{E}$ is Earth's equatorial radius.

One can also construct the feasible region of satellite positions, which
corresponds to the intersection of all interior ($\alpha_0 > 0$) or exterior
($\alpha_0 < 0$) points of each reference station's feasibility cone.  In the
special case of $\alpha_0 = 0$, feasible region construction becomes a nonlinear
optimization problem with linear constraints formulated by enforcing
$\vb{r} \in \mathbb{R}^3$ to be above the tangent plane at each reference
station.  The general case is similar but with nonlinear constraints---those in
\eqref{eq:nlrMin}.

One might assume that the reference stations not detecting interference could be
used to further constrain the feasible region. However, this assumption does not
hold when the antenna gain pattern of the interference source is unknown. For
example, if the interference source is equipped with a narrow-beam antenna, many
reference stations may not observe the interference even if the satellite
satisfies the elevation mask constraint.

Fig.~\ref{fig:region} illustrates the positions of all publicly tracked objects
during the high-power interference burst on day 160 of 2021. Its right figure
shows the subset of satellites satisfying $\alpha_0 = 0$, excluding known debris
and rocket bodies. The interior of the quasi-conical red surface represents the
feasible region in which the interference satellite could have been
located. Table \ref{tbl:elMask160} shows $\zeta^*$ for various elevation mask
angles for this event, as well as the number of satellites that satisfy the
elevation mask, e.g., there are 201 satellites that satisfy the 0$^\circ$
elevation mask.
\ifpreprint
	\begin{figure}[t]
	\centering
	\begin{minipage}[b]{0.44\textwidth}
		\centering
		\includegraphics[width=\linewidth]{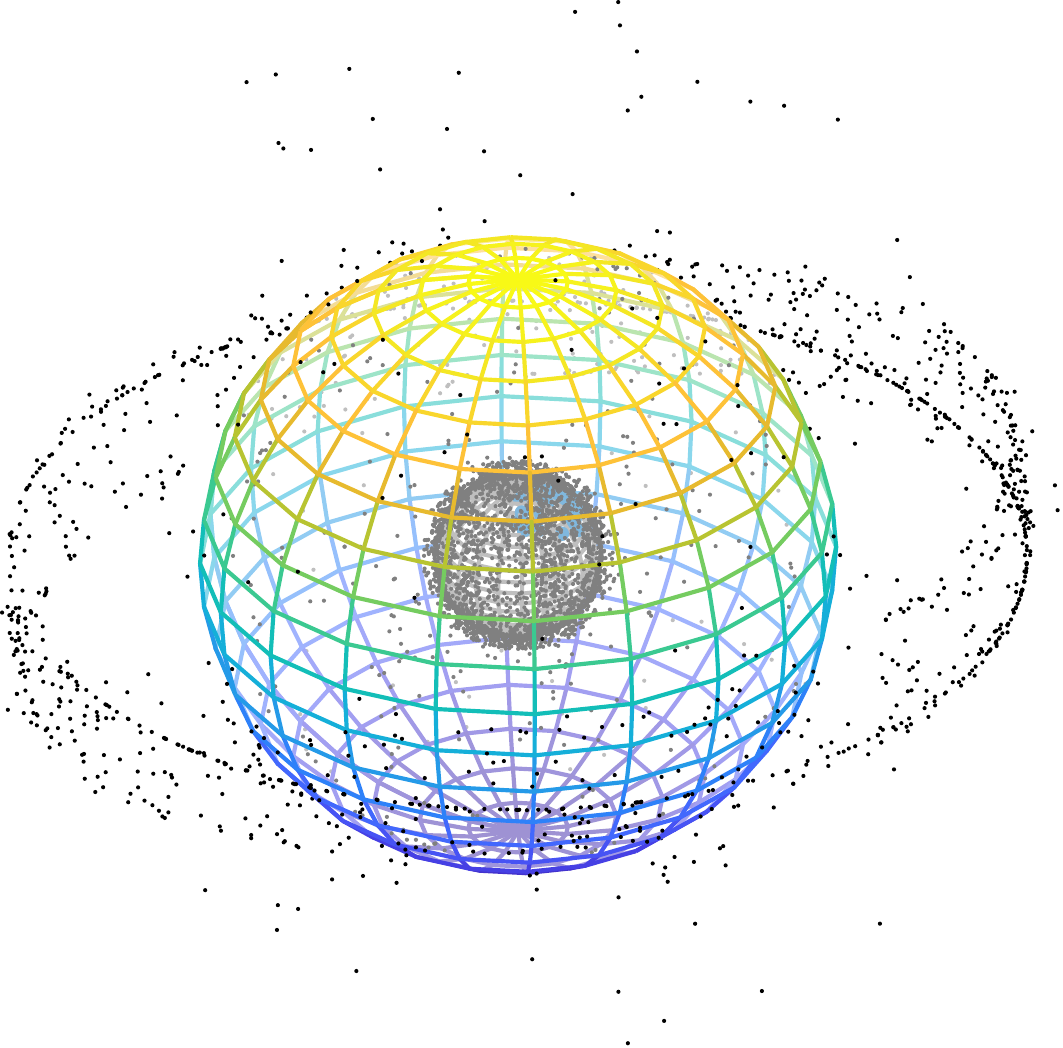}
	\end{minipage}
	\hspace{10mm}
	\begin{minipage}[b]{0.44\textwidth}
		\centering
		\includegraphics[width=\linewidth]{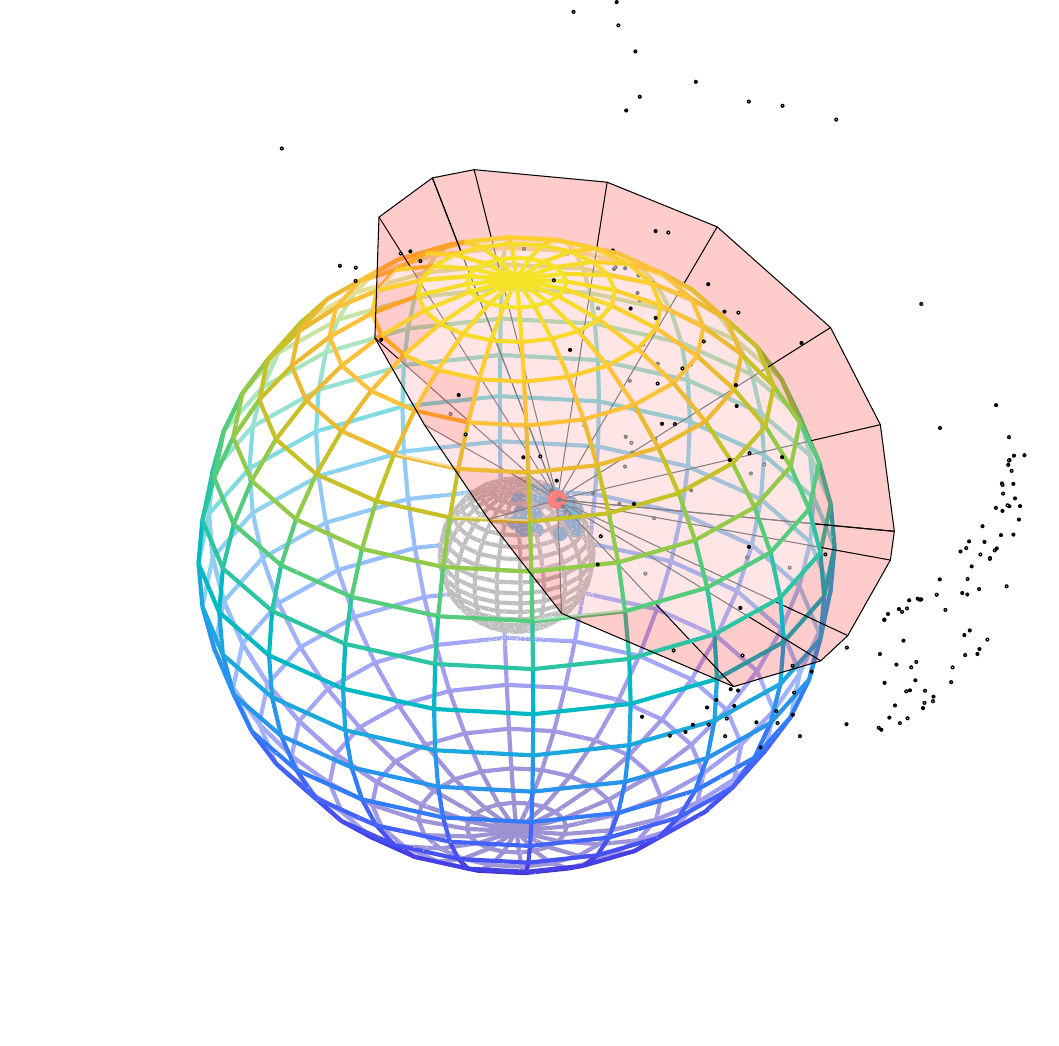}
	\end{minipage}
	\caption{Left: The position of all tracked objects during the high-power
		interference burst on day 160 year 2021.  Right: The position of all
		satellites that satisfy a 0$^\circ$ elevation mask, excluding debris and
		rocket bodies. The feasible region in which the interference source could
		have been positioned is interior to the red surface.  For reference, the
		colored spherical shell corresponds to medium Earth orbit (20,000 km
		altitude).}
	\label{fig:region}
\end{figure}
\else
	\begin{figure}[t]
	  \centering
	  \begin{minipage}[b]{0.4\textwidth}
	    \centering
	    \includegraphics[width=\linewidth]{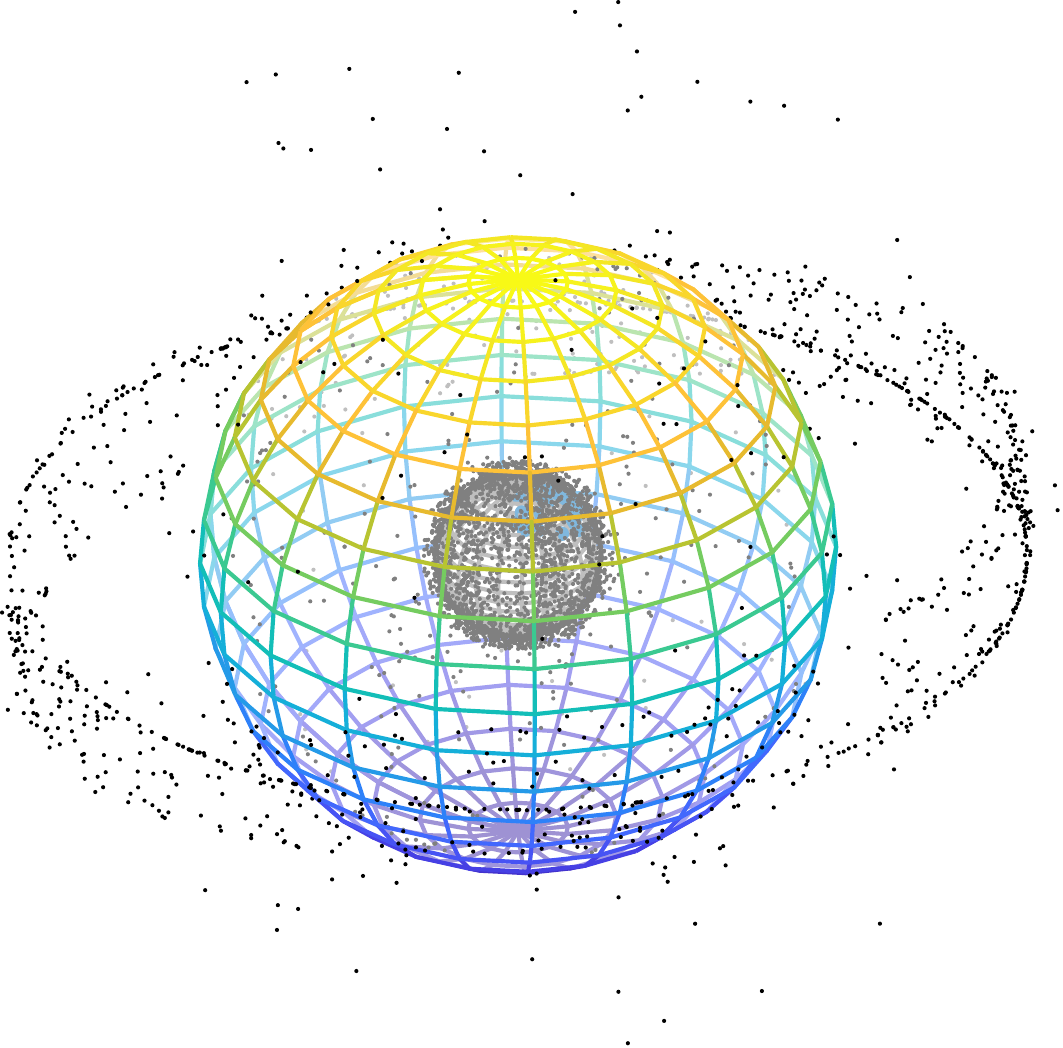}
	  \end{minipage}
	  \hspace{10mm}
	  \begin{minipage}[b]{0.4\textwidth}
	    \centering
	    \includegraphics[width=\linewidth]{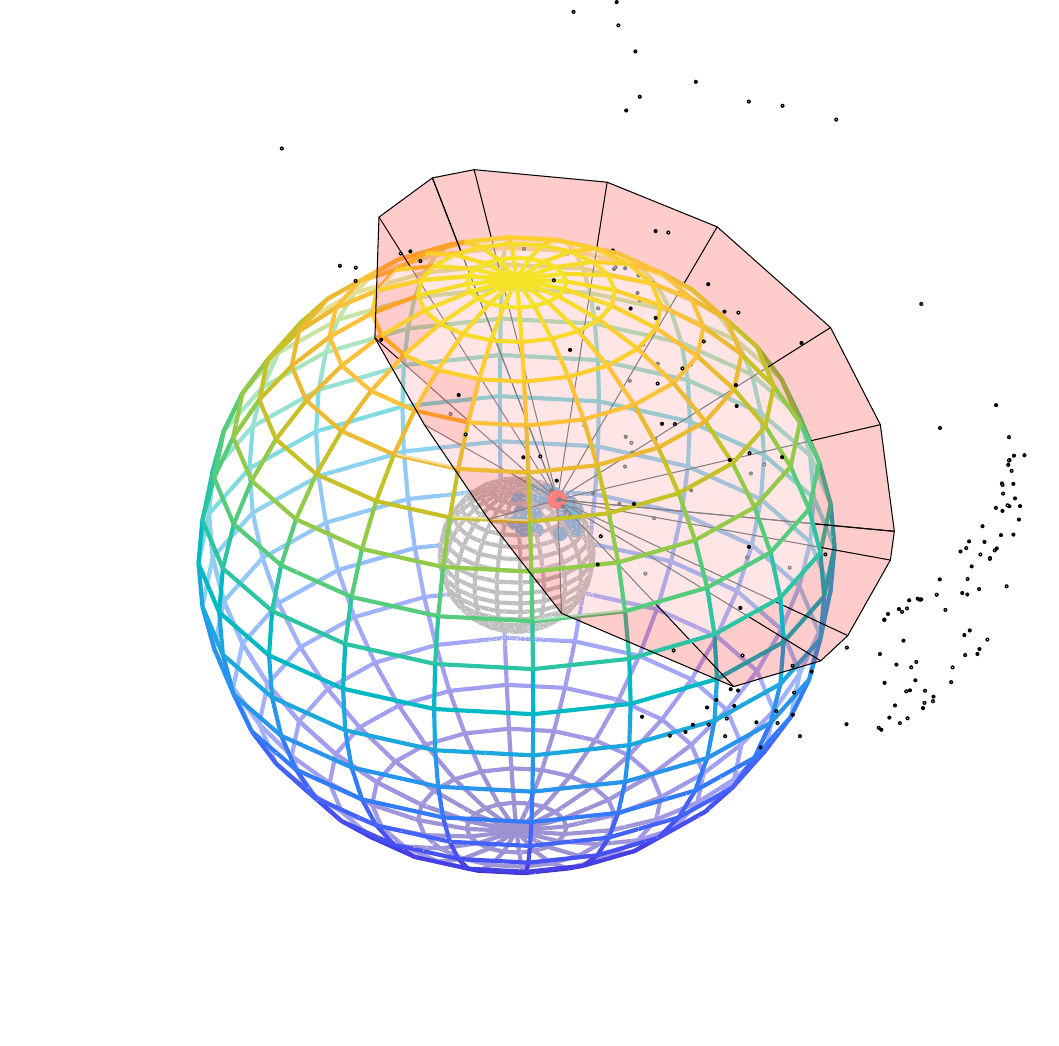}
	  \end{minipage}
	  \caption{Left: The position of all tracked objects during the high-power
	    interference burst on day 160 year 2021.  Right: The position of all
	    satellites that satisfy a 0$^\circ$ elevation mask, excluding debris and
	    rocket bodies. The feasible region in which the interference source could
	    have been positioned is interior to the red surface.  For reference, the
	    colored spherical shell corresponds to medium Earth orbit (20,000 km
	    altitude).}
	  \label{fig:region}
	\end{figure}
\fi

\begin{table}[t]
  \caption{The minimum altitude and number of satellites that satisfy various elevation masks for the event on day 160 of 2021}
  \centering
  \normalsize
  \begin{tabular}{ c | c c c c c c c c   } 
    \toprule
    $\alpha_0$ & -5$^\circ$ & -3$^\circ$ & -1$^\circ$ & 0$^\circ$ & 1$^\circ$ & 3$^\circ$ & 5$^\circ$ & 10$^\circ$  \\ 
    \hline
    $\zeta^*$ [km] & 800 & 957 & 1,120 & 1,212 & 1,290 & 1,468 & 1,655 & 2,173  \\ \hline
    Number of valid satellites &   381 &   303  &   229  &   201  &   176  &   145  &   113  &    87 \\ 
    \bottomrule
  \end{tabular}
  \label{tbl:elMask160}
\end{table}
If one assumes that a single satellite is responsible for the observed
interference events, the list of candidate satellites can be further narrowed by
eliminating those that do not satisfy the elevation mask across multiple events.
As shown later, however, the single-satellite assumption does not hold for this
paper's transient interference.

\section{Multi-Hypothesis CNR-Based Interference Source Identification}
\label{sec:CNR_SatID}
This section develops a more rigorous CNR-based technique to associate
measurements to a satellite than the simple elevation-mask-based method of the
previous section.

\subsection{Generalized Likelihood Ratio Test}
A binary Neyman-Pearson hypothesis testing framework provides the optimal test
for simple cases in which the relevant probability density functions under both
the null ($H_0$) and alternative ($H_1$) hypotheses are fully specified.  The
test is optimal in the sense of maximizing the probability of detection for a
fixed probability of false alarm.  In real-world scenarios, however, some
parameters are often unknown, resulting in composite tests.  If the unknown
parameters' \emph{a~priori} probability distributions are also unknown, then no
known optimal test exists \citep{vtrees2001dem}.  Such conditions apply to this
paper's attempt to identify the source of space-based interference using
received-power measurements because parameters such as the source's transmit
power are unknown.

The Generalized Likelihood Ratio Test (GLRT) is often employed for composite
detection problems with unknown prior distributions.  It amounts to a two-step
procedure.  First, the unknown parameter vector $\vb{x}$ is estimated from the
measurements $\vb{z}$ via maximum likelihood estimation under each hypothesis.
Second, these estimates are used to evaluate likelihoods as if they were the
correct values of $\vb{x}$ under the respective hypotheses
\citep{vtrees2001dem}.
In the current context, the hypothesis test is M-ary rather than binary, with
$H_s$ being the hypothesis that satellite $s \in \mathcal{S}$ is the
interference source. The vector $\vb{x}_s$ contains parameters assumed unknown
\emph{a priori} that are used to model the candidate satellite's transmit power,
antenna gain pattern, and pointing direction (beam vector), or some subset of
these.

\subsection{Simulation Study}
The performance of a multi-hypothesis satellite association technique may be
illustrated by a simulation study.  This section's study applies an M-ary
version of the GLRT not to the interference source identification problem itself
but to a closely related problem: identifying which satellite is the source of a
GNSS signal whose CNR measurements are taken from a network of reference
stations.  This related problem is chosen because (1) it yields insights
pertinent to the interference identification problem, and (2) the technique
developed and demonstrated in simulation can be tested on real-world CNR
measurements from the IGS network to demonstrate identification of a known
source GNSS satellite.

\begin{figure}[t]
  \centering
  \begin{minipage}[b]{0.44\textwidth}
    \centering
    \includegraphics[width=\linewidth]{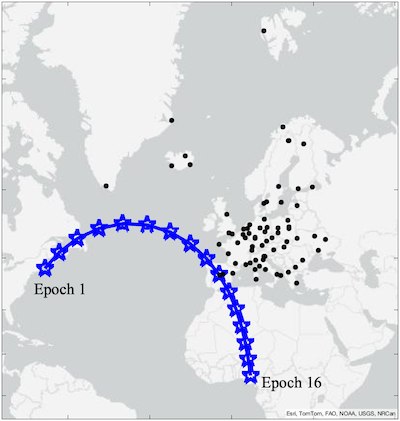}
  \end{minipage}
  \hspace{8mm}
  \begin{minipage}[b]{0.47\textwidth}
    \centering
    \includegraphics[width=\linewidth]{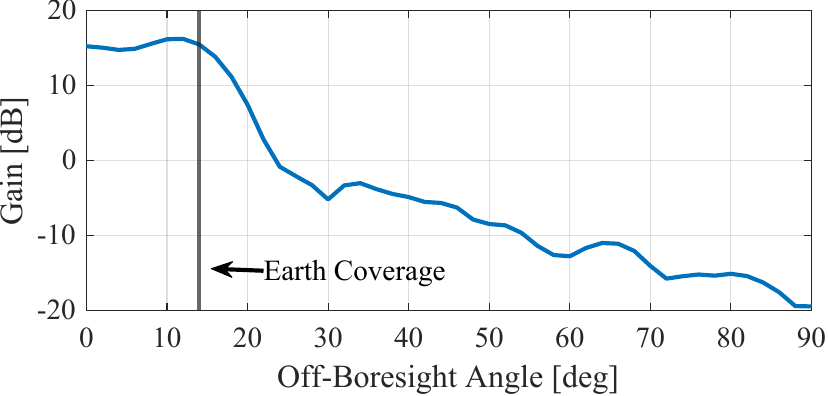}
    \vspace{2mm}
    \includegraphics[width=\linewidth]{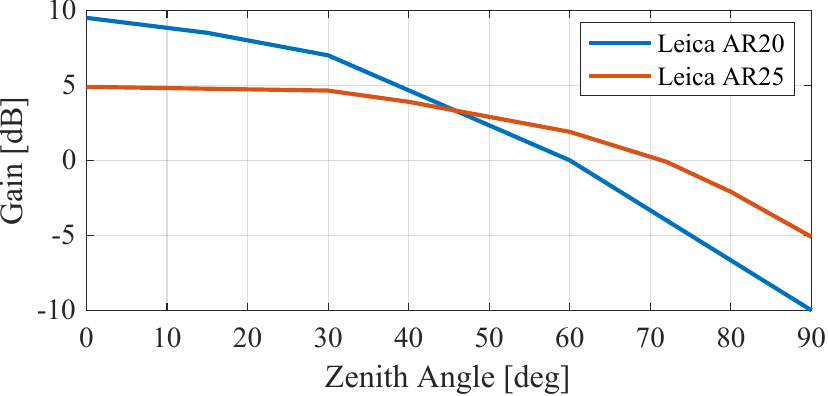}	
  \end{minipage}
  \caption{Simulation study setup.  Left: The geometry of the true source GNSS
    satellite over 16 epochs at 15-minute intervals, with terrestrial reference
    stations shown as black dots.  Right: Transmitter (top) and receiver
    (bottom) gain patterns. The receiver gain patterns are from a model provided
    by the manufacturer and verified empirically.}
  \label{fig:simulation}
\end{figure}

For the $i$th station and $j$th GNSS signal at epoch $k$, let $z_{ij}[k]$ be an
interference-free CNR measurement, modeled as in \eqref{eq:H0meas}.  The
measurement noise variance $\sigma_{ij}^{2}[k] = f(\theta_\text{R}^{ij}[k])$ is
modeled as a nonlinear function of the received zenith angle, with
$\sigma_{ij} = 0.25$~dB at zenith ($\theta_\text{R}^{ij}[k]=0$) and
$\sigma_{ij} = 1.25$~dB at the horizon ($\theta_\text{R}^{ij}[k]=90^\circ$).
Both receiver and transmitter antenna gain patterns are modeled as azimuthally
symmetric---$G_\text{R}^i(\theta, \phi) = G_\text{R}^i(\theta)$,
$G_\text{T}^j(\theta, \phi) = G_\text{T}^j(\theta)$---which is approximately
true for both GNSS receiver and transmitter antennas.  Fig.~\ref{fig:simulation}
shows the assumed reference station locations (black dots) and the position of
an exemplar GNSS satellite for 16 time epochs spaced by 15-minute intervals.  It
also shows the corresponding transmitter and receiver gain patterns. In this
study, all transmitter antennas are nadir-pointing, and all reference stations
are assumed to utilize either Leica AR20 or AR25 antennas.

At epoch $k$, let $\mathcal{I}_j[k] \subseteq \mathcal{I}$ be the set of
participating stations tracking the $j$th GNSS signal, and $\mathcal{S}_\text{e}[k]$ be
the set of candidate source satellites that have all passed elevation mask
vetting. Let the measurement vector $\vb{z}_j[k]$ contain ordered CNR
measurements of signal $j \in \mathcal{J}$ from those receivers tracking the
signal at epoch $k$, i.e., measurements from the set
$\{z_{ij}[k] \mid i \in \mathcal{I}_j[k] \}$.

A GLRT-based association test is used to assess whether a candidate satellite
$s \in \mathcal{S}_\text{e}[k]$ is consistent with $\vb{z}_j[k]$ given varying degrees of
prior knowledge about the satellite's transmitter.  The unknown parameter vector
$\vb{x}_s[k]$ may include any combination of transmit power $P_s[k]$, beamwidth
$\beta_s[k]$, and beam vector $\vb{v}_s[k] = [\theta_s[k], ~\phi_s[k]]\T$, where
$\theta_s[k]$ and $\phi_s[k]$ are the off-boresight and azimuth angle relative
to the nadir vector and a reference azimuth vector in the satellite frame.  In
the fully unknown case, $\vb{x}_s$ includes all three parameters:
\begin{align}
  \vb{x}_s[k] = [ P_s[k] \, , \, \beta_s[k] \, ,\, \vb{v}_s\T[k]]\T
\end{align}

Under hypothesis $H_s$, satellite $s \in \mathcal{S}_\text{e}[k]$ is the source
of signal $j \in \mathcal{J}$, in which case the measurement vector
$\vb{z}_j[k]$ is distributed as
\begin{align}
  \label{eq:Hsdist}
  H_s:\quad \vb{z}_j[k] \sim \mathcal{N}\left(\vbb{z}_j[k, \vb{x}_s],~ R_j[k, \vb{x}_s]\right),
  \quad s \in \mathcal{S}_\text{e}[k]
\end{align}
Here, $\vbb{z}_j[k, \vb{x}_s]$ is the vector of expected CNR measurements for
signal $j\in \mathcal{J}$ emanating from satellite $s \in \mathcal{S}_\text{e}[k]$, and
$R_j[k,\vb{x}_s]$ is a square diagonal matrix with diagonal elements
$\sigma_{ij}^{2}[k]$ for $i \in \mathcal{I}_j[k]$, where the angle
$\theta_\text{R}^{ij}[k]$, on which $\sigma_{ij(s)}^{2}[k]$ depends, is
calculated assuming signal $j$ comes from satellite $s$.

This association problem amounts to a multi-hypothesis composite general
Gaussian problem \citep{vtrees2001dem}.  Hypotheses may be compared based on
their negative log likelihood function evaluated at the maximum-likelihood value
for the parameter vector $\vb{x}_s$, which, for the model in \eqref{eq:Hsdist},
leads to a cost function of the form
\begin{equation}
  \label{eq:costFcn}
  J_j[k, \vb{x}^*_s] = \left(\vb{z}_j[k] - \vbb{z}_j[k,\vb{x}^*_s] \right)\T
  \left(R_j[k,\vb{x}^*_s]\right)^{-1} \left(\vb{z}_j[k] -
    \vbb{z}_j[k,\vb{x}^*_s] \right), \quad s \in \mathcal{S}_\text{e}[k]
\end{equation}
where $\vb{x}^*_s$ is the value of $\vb{x}_s$ that minimizes $J_j[k, \vb{x}_s]$.
Thus, for $a,s \in \mathcal{S}_\text{e}[k]$, if
$J_j[k, \vb{x}^*_a] < J_j[k, \vb{x}^*_s]$, then $H_a$ is more likely than $H_s$.
Since $\mathcal{S}_\text{e}[k]$ is assumed to include the true source for signal $j$,
then for the case of hypotheses with equally likely priors, the probability that
$H_s$ is true is given by \citep{y_barshalom01_tan}
\begin{equation}
  \label{eq:probBS}
  P(s | \vb{z}_j[k]) = \frac{|R_j[k,\vb{x}^*_s]|^{-1/2}\exp\left(-\tfrac{1}{2}J_j[k, \vb{x}^*_s]
    \right)}{\displaystyle\sum_{a \in \mathcal{S}_\text{e}[k]} |R_j[k,\vb{x}^*_a]|^{-1/2}
    \exp\left(-\tfrac{1}{2}J_j[k, \vb{x}^*_a] \right)}, \quad s \in \mathcal{S}_\text{e}[k]
\end{equation}
Based on this expression, one could choose the highest-probability satellite, or
the top $N$ most probable, or all those for which $P(s | \vb{z}_j[k])$ exceeds a
threshold.  Nearly equivalent to the latter approach, one may retain all
$s \in \mathcal{S}_\text{e}[k]$ for which $J_j[k, \vb{x}^*_s] \leq \nu_{j}[k]$
for a given threshold $\nu_{j}[k]$.  A principled choice for $\nu_{j}[k]$ can be
obtained by limiting the probability
$P\left(J_j[k, \vb{x}^*_u] > \nu_{j}[k]\right)$ to a specified size, where
$u \in \mathcal{S}_\text{e}$ is the true source satellite.  It is
straightforward to show that if $\vb{x}^*_u$ is an accurate estimate then
$J_j[k, \vb{x}^*_u]$ is distributed as a chi-square random variable with
$|\mathcal{I}_j[k]|$ degrees of freedom:
\begin{equation}
  \label{eq:chi2J}
  J_j[k, \vb{x}^*_u] \sim \chi^2_{|\mathcal{I}_j[k]|}  
\end{equation}
From this one can readily calculate $\nu_{j}[k]$ to ensure
$P\left(J_j[k, \vb{x}^*_u] > \nu_{j}[k]\right)$ is below a chosen value.

\subsubsection{Simulation}
1000 Monte Carlo trials were conducted at each time epoch depicted in
Fig.~\ref{fig:simulation}. A trajectory for GPS PRN 15 from February 2020 was
chosen for the true source satellite $s$. At each epoch, (1) a subset
$\mathcal{S}_\text{e}[k]$ of candidate satellites was generated by applying an
elevation mask $\alpha_0 = 0$; (2) the subset $\mathcal{I}_j[k]$ was determined
for signal $j$; (3) the noise-free CNR$_{ij}$ at each reference station
$i \in \mathcal{I}_j[k]$ was computed based on the true parameter vector
$\vb{x}_s$ under the model \eqref{eq:CNRij1}; (4) a realization of the
measurement $z_{ij}[k]$ was generated according to \eqref{eq:H0meas} assuming
$\sigma_{ij}^{2}[k] = f(\theta_\text{R}^{ij}[k])$ for each
$i \in \mathcal{I}_j[k]$.

Costs $J_j[k, \vb{x}^*_s]$ were calculated for all $s \in \mathcal{S}_\text{e}[k]$, with
$\vb{x}_s$ estimated under five scenarios reflecting varying levels of \emph{a
  priori} knowledge of the transmitter: (S1) all parameters known; (S2) unknown
transmit power $P_s$; (S3) unknown $P_s$ and beam width $\beta_s$; (S4) unknown
$P_s$ and beam vector $\vb{v}_s$; and (S5) unknown $P_s$, $\beta_s$, and
$\vb{v}_s$.  The maximum likelihood (minimum cost) estimates $\beta_s^*$, and
$\vb{v}_s^*$ were found via grid search, while $P^*_s$ was taken as the mean of
the elements of the difference $\vb{z}_j[k] - \vbb{z}_j[k,\vbt{x}_s^*]$, with
$\vbt{x}^*_s = [ P_s = 0 \, , \, \beta^*_s \, ,\, \vb{v}^{*\mathsf{T}}_s]\T$.

\subsubsection{Results and Discussion}
\begin{figure}[t]
  \centering
  \begin{minipage}[b]{0.49\textwidth}
    \centering
    \includegraphics[width=\linewidth]{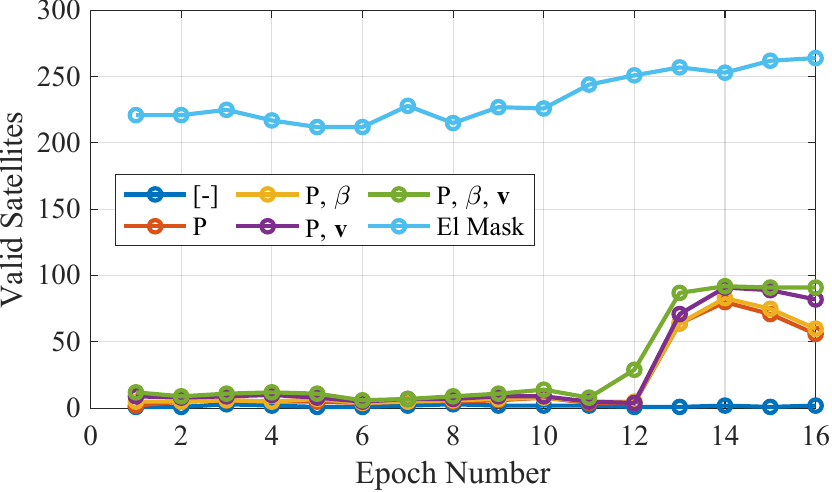}
  \end{minipage}
  \hfill
  \begin{minipage}[b]{0.49\textwidth}
    \centering
    \includegraphics[width=\linewidth]{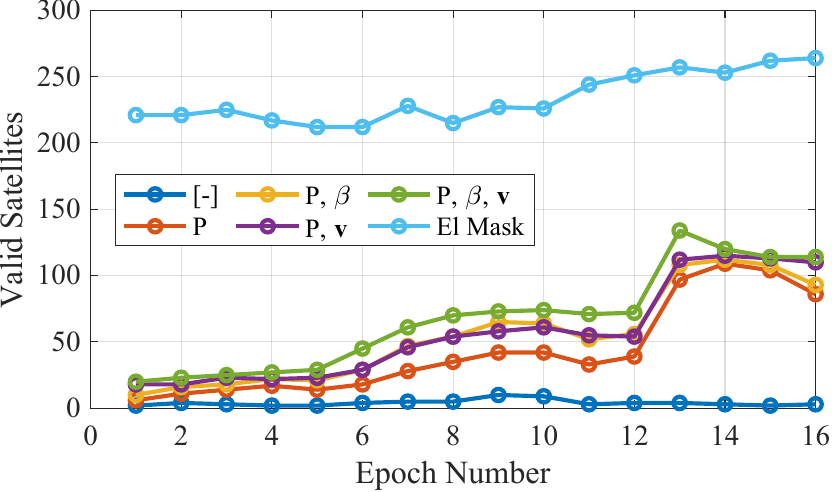}
  \end{minipage}
  \caption{The number of satellites that, for at least 5\% of the Monte Carlo
    trials, remained viable candidates after association testing with
    $P\left(J_j[k, \vb{x}^*_s] > \nu_{j}[k]\right)< 10^{-3}$.  The legends
    indicate which parameters are assumed unknown. The number of satellites
    satisfying all elevation masks at each epoch, $|\mathcal{S}_\text{e}[k]|$, is shown
    in light blue.  Left: All reference stations modeled with Leica AR20
    antennas.  Right: All reference stations modeled with Leica AR25 antennas.}
  \label{fig:simulationResults}
\end{figure}

Fig.~\ref{fig:simulationResults} shows the number of satellites that, for at
least 5\% of the Monte Carlo trials, were considered viable, i.e., had costs
$J_j[k, \vb{x}^*_s]$ below $\nu_{j}[k]$ when
$P\left(J_j[k, \vb{x}^*_u] > \nu_{j}[k]\right)$ was limited to under $10^{-3}$.
As shown, the analysis was performed across various epochs, transmitter
scenarios, and receiver antenna types.  The results reveal that the GLRT-based
multi-hypothesis technique is far superior to multi-station elevation masking in
narrowing the number of candidate satellites.  When transmitter power,
beamwidth, and beam vector are known \emph{a priori}, only the true satellite
and a small number of others are consistent with the data at any given epoch, as
shown by the dark blue trace along the bottom of both panels in
Fig.~\ref{fig:simulationResults}.  Even when none of these parameters is known
\emph{a priori} and so they must be estimated from the network data (green
traces in Fig.~\ref{fig:simulationResults}), the method offers useful
discrimination.  But it is clear that, at the geometries arising during many of
the epochs, the CNR-based association method alone would not be sufficient for
confident single-epoch satellite identification when there are unknown
parameters.

The two panels in Fig.~\ref{fig:simulationResults} also show that the receiver
antenna gain pattern substantially influences the test's discrimination
power. Specifically, the Leica AR20 (left panel) exhibits a more pronounced
variation in gain over the zenith angle compared to the Leica AR25 (right
panel), thereby enhancing the sensitivity of measurements to changes in zenith
angle and facilitating the exclusion of false associations.

The impact of transmitter satellite geometry on correct association is also
evident.  Note how the number of viable satellites increases when the true
satellite approaches the equator during the last few epochs.
Fig.~\ref{fig:simulationEpochs} offers an explanation: Compared to epoch
$k = 1$, which is far from the equator and saw relatively few false
associations, at epoch $k = 16$ many more satellites are falsely associated
because a whole band of geostationary satellites are also consistent with the
data when any transmitter parameters are unknown \emph{a~priori}.

\ifpreprint
	\begin{figure}[t]
	  \centering
	  \includegraphics[width=.85\linewidth]{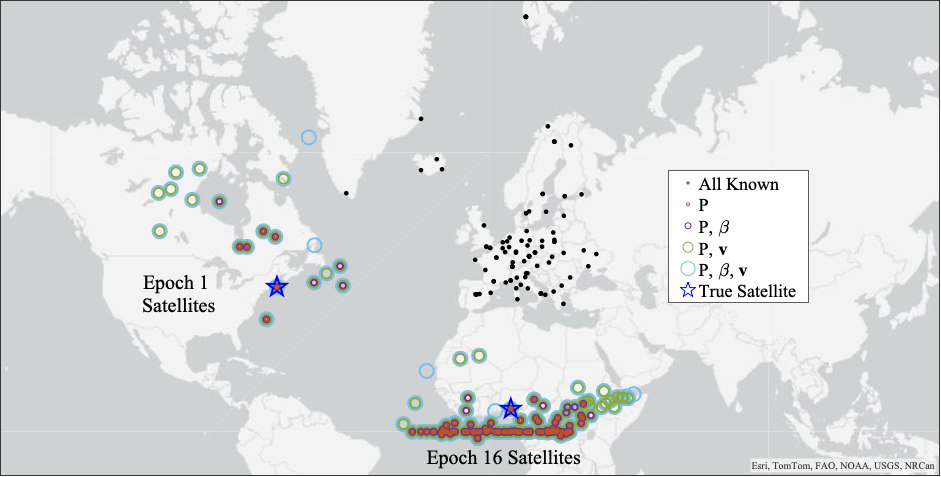}
	  \caption{Satellites that remained viable candidates on more than 5\% of Monte
	    Carlo trials for epoch $k = 1$ and epoch $k = 16$ when the reference
	    stations are modeled as using the Leica AR25 antenna.  The legend gives
	    symbols for the true satellite and for valid candidates under scenarios with
	    an increasing number of unknown transmitter parameters.}
	  \label{fig:simulationEpochs}
	\end{figure}
\else
	\begin{figure}[t]
		\centering
		\includegraphics[width=.8\linewidth]{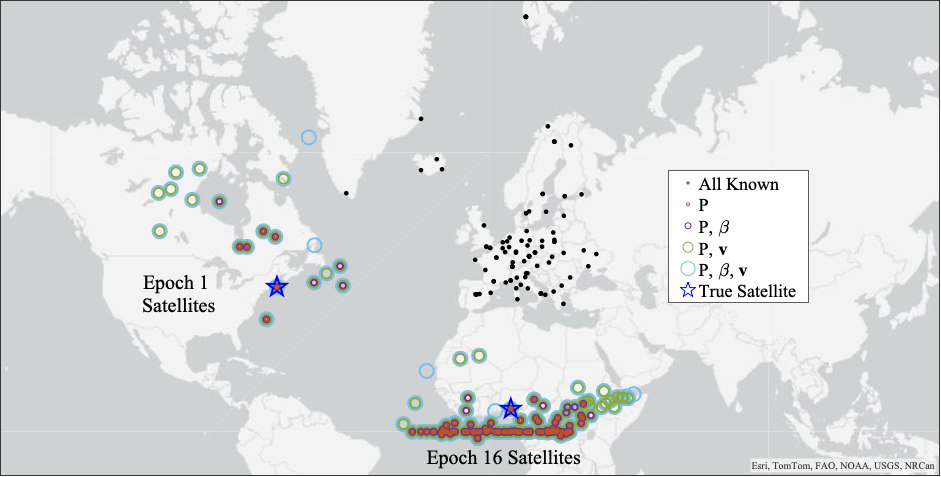}
		\caption{Satellites that remained viable candidates on more than 5\% of Monte
			Carlo trials for epoch $k = 1$ and epoch $k = 16$ when the reference
			stations are modeled as using the Leica AR25 antenna.  The legend gives
			symbols for the true satellite and for valid candidates under scenarios with
			an increasing number of unknown transmitter parameters.}
		\label{fig:simulationEpochs}
	\end{figure}
\fi

\subsection{GNSS Satellite Identification with Real CNR Data}
\begin{figure}[t]
  \centering
  \begin{minipage}[b]{0.48\textwidth}
    \centering
    \includegraphics[width=\linewidth]{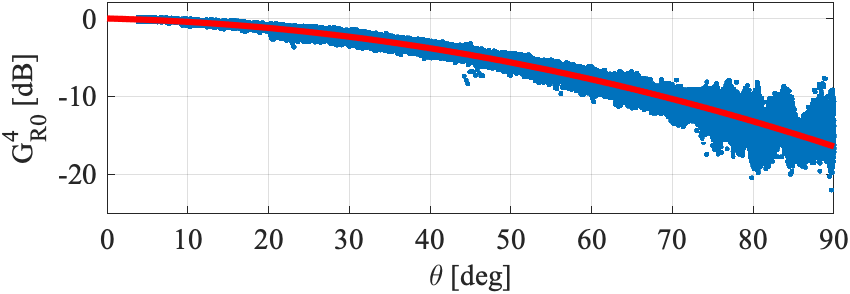}
  \end{minipage}
  \hfill
  \begin{minipage}[b]{0.48\textwidth}
    \centering
    \includegraphics[width=\linewidth]{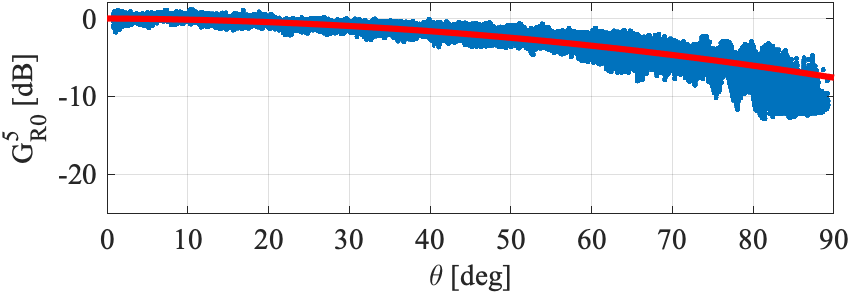}
  \end{minipage}
  \caption{Normalized estimates of $G_{\text{R}0}^i(\theta)$ for stations MEDI
    (Leica AR20, $i = 4$) and NICO (Leica AR25, $i = 5$). The blue dots
    represent collected values of $\text{CNR}_{ij}~-~G_\text{T}^j~-~L_{ij}$,
    while the red line indicates a weighted third-order polynomial fit, which
    becomes $G_{\text{R}0}^i(\theta)$, all normalized to
    $G_{\text{R}0}^i(0)=0$.  As expected, measurement noise increases
    significantly as the elevation angle decreases.}
  \label{fig:Grx}
\end{figure}

The GNSS satellite identification technique developed here and demonstrated by
simulation was also tested using real CNR measurements taken from reference
stations in $\mathcal{I}$.  Let $\mathcal{J}_\text{g} \subset \mathcal{J}$ be
the subset of GPS L1 C/A signals.  Based on historical data from each site
$i \in \mathcal{I}$ and on the known azimuthally symmetric antenna gain pattern
$G^j_\text{T}(\theta)$ and transmit power $P^j_\text{T}$ for
$j \in \mathcal{J}_\text{g}$ from a select number of GPS satellites
$\mathcal{S}_\text{g} \subset \mathcal{S}_\text{e}$, empirical estimates of the combined
receiver antenna gain pattern and receiver noise floor
\begin{equation}
  \label{eq:combinedGRNo}
  G_{\text{R}0}^i(\theta) =  G_\text{R}^i(\theta) - N^i_0
\end{equation}
were obtained for all $i\in \mathcal{I}$ across zenith angles
$\theta \in [0, 90^\circ]$, assuming azimuthal symmetry.  Accurate estimates of
$G^j_\text{T}(\theta)$ for $j \in \mathcal{J}_\text{g}$ and
$s \in \mathcal{S}_\text{g}$ are available in public documents
\citep{marquis2015gps}, and corresponding estimates for $P^j_\text{T}$ were
obtained by observation with a calibrated high-gain antenna.  Normalized
estimates of $G_{\text{R}0}^i(\theta)$ for stations MEDI (Leica AR20) and NICO
(Leica AR25) are shown in Fig.~\ref{fig:Grx}.

The GLRT-based multi-hypothesis association test was then applied based on
measured CNR$_{ij}$ values for $i \in \mathcal{I}$ and
$j \in \mathcal{J}_\text{g}$ for $s \in \mathcal{S}_\text{g}$, according to
\eqref{eq:H0meas} with $\sigma_{ij}^{2} = f_i(\theta_\text{R}^{ij})$.  The
site-specific function $f_i$ was determined by the zenith-angle-dependent spread
in the empirical data underlying each station's antenna model (e.g., the blue
dots in Fig. \ref{fig:Grx}).  The method was tested under full \emph{a priori}
knowledge of the transmit power, beamwidth, and beam vector of
$s \in \mathcal{S}_\text{g}$, and under a case where the transmit power $P_s$
was unknown.  The results of this empirical study were qualitatively consistent
with those of the simulation study: the true $s \in \mathcal{S}_\text{g}$ was
often uniquely identifiable in a single epoch when its transmitter parameters
were all known \emph{a priori}, but false associations increased in
geometry-dependent ways when $P_s$ had to be estimated from the data.

\subsection{Implications for CINR-Based Interference Source Identification}
\label{subsec:implicationsForCINRSourceID}
Based on CNR and CINR measurements alone, interference source identification is
more challenging than GNSS satellite identification in two key respects.  First,
the transmitter gain pattern may be wholly unknown, not just unknown in
beamwidth $\beta_s$ as in the simulation study.  Second, and more importantly,
the receiver antenna gain pattern $G_\text{R}^i(\theta)$ and noise floor $N^i_0$
must be known independently, rather than combined as in \eqref{eq:combinedGRNo},
because the model for CINR$_{ij}$ given in \eqref{eq:CINRij} involves an
isolated value of $\tilde{N}^i_0$.  Yet it is not possible to independently
estimate $G_\text{R}^i(\theta)$ and $N^i_0$ for $i \in \mathcal{I}$ strictly
from historical CNR measurements because CNR is a linear combination of the two
quantities, as evident in \eqref{eq:CNRij1} and \eqref{eq:PijR}.  Separate
estimates of $N^i_0$ can be obtained for $\mathcal{I}_\text{a} \in \mathcal{I}$,
the subset of reference stations for which an accurate antenna gain pattern
$G_\text{R}^i(\theta)$ is available, e.g., one provided by the manufacturer.
But the authors of the current paper found $\mathcal{I}_\text{a}$ was limited to
stations with Leica AR20 or AR25 antennas. For all other stations (all
$i \in \mathcal{I} \setminus \mathcal{I}_\text{a}$), either
$G_\text{R}^i(\theta)$ was not available or the available model did not match
the shape of the empirically derived combined model $G_{\text{R}0}^i(\theta)$.
(Note that, whereas the IGS provides extremely accurate models of antenna phase
center variation as a function of $\theta$ and $\phi$ \citep{krzan2020antenna},
it provides no such models for antenna gain.)  Importantly, the geographic
distribution of $\mathcal{I}_\text{a}$ is not as extensive as that shown in
Fig.~\ref{fig:simulation}, leading to weaker observability of the unknown
transmitter parameters, and thus weaker discrimination of candidate satellites.

In view of these challenges, one can expect the current performance of CNR-based
multi-hypothesis association for interference source identification to fall
somewhere between the light blue and green curves shown in
Fig.~\ref{fig:simulationResults}, tending towards the light blue curve when the
unknown parameters multiply. It becomes clear that, to confidently identify a
GNSS interference source, more powerful techniques are required.  Nonetheless,
CNR-based multi-hypothesis association could offer useful prior probabilities
for other interference source identification techniques if accurate gain
patterns $G_\text{R}^i(\theta)$ were available for more stations in
$\mathcal{I}$.  Moreover, many stations provide CNR measurements only as
integer-quantized values. Reporting these measurements at a higher resolution
would significantly enhance measurement accuracy.

Assuming these desiderata for IGS stations will eventually be fulfilled, the
mapping from the GNSS satellite identification problem to the interference
source identification problem is briefly outlined.  First, supposing that
$G_\text{R}^i(\theta)$ is known, and leveraging the known transmit gain pattern
$G^j_\text{T}(\theta)$ and transmit power $P^j_\text{T}$ for
$j \in \mathcal{J}_\text{g}$, $N_0^i$ is estimated from historical CNR data for
all $i\in \mathcal{I}$ so that the models \eqref{eq:CINRij} and
\eqref{eq:H1meas} can be applied.  Then, assuming that at epoch $k$ there is an
interference event, let $\mathcal{I}_\text{d}[k] \subset \mathcal{I}$ be the
subset of stations that detect the event, and let the vector $\vb{z}[k]$ contain
ordered station-specific detection statistics at epoch $k$, i.e., values from
the set $\{\Lambda_i[k] \mid i \in \mathcal{I}_\text{d}[k] \}$.  For every
satellite $s \in S[k]$ and for an assumed parameter vector $\vb{x}_s$, one can
develop a hypothesis model for $\vb{z}[k]$ like the one in \eqref{eq:Hsdist} and
find the subset of $S[k]$ consistent with $\vb{z}[k]$, just as with the GNSS
satellite identification problem.  Let the resulting probabilities, analogous to
$P(s | \vb{z}_j[k])$ in \eqref{eq:probBS}, be denoted $P(s | \vb{z}[k])$.  The
following section shows how these probabilities can act as prior probabilities
for a more powerful identification technique.

\section{TDOA-Based Interference Source Identification}
This section presents a satellite association framework based on TDOA
measurements, together with a TDOA analysis based on recently captured raw
in-phase and quadrature (IQ) samples from two spatially separated receivers in
Europe during a continent-scale interference event in February 2026.

\subsection{TDOA Measurement Models}
Consider two receivers at fixed locations $\vb{g}_1$ and $\vb{g}_2$ in the ECEF
reference frame. Let $\mathcal{K}=\{1,\dots, K\}$ be the set of TDOA measurement
epochs over an observation interval, $\mathcal{S}_\text{e}$ be the set of candidate
satellites that satisfy elevation mask vetting for all $k~\in~\mathcal{K}$, and
$\vb{r}_{s}[k]$ be the ECEF position of satellite $s~\in~\mathcal{S}_\text{e}$ at epoch
$k$.  For model simplicity, $\vb{r}_{s}[k]$ is assumed to remain constant over
the signal's time-of-flight for each measurement.  Denoting by
$\rho_{is}[k]=\|\vb{g}_i - \vb{r}_{s}[k]\|$ the range between receiver
$i \in \{1,2\}$ and satellite $s \in \mathcal{S}_\text{e}$, the modeled TDOA for $s$ is
the range difference between the two receivers:
$\bar{y}_s[k]=\rho_{2s}[k]-\rho_{1s}[k]$.  Assume that $u \in \mathcal{S}_\text{e}$ is
the true interference source.  Then the actual TDOA measurement is modeled as
\begin{equation}
  \label{eq:tdoaModel_s}
  y[k] = \bar{y}_u[k] + w[k], \quad k \in \mathcal{K}
\end{equation}
where the AWGN noise term $w[k] \sim \mathcal{N}(0, \sigma^{2})$ accounts for
the aggregate effects of thermal noise and quantization error from
nearest-neighbor sampling.  Epoch-wise measurements are stacked as
$\vb{y}=[y[1],\dots,y[K]]\T$, and modeled values for $s\in\mathcal{S}_\text{e}$
as $\bar{\vb{y}}_s=[\bar{y}_s[1],\dots,\bar{y}_s[K]]\T$.

The location $\vb{r}_{s}[k]$ of each $s \in \mathcal{S}_\text{e}$ is not known exactly
but can be approximated with the public TLE data available for $s$.  Let
$\vbt{r}_s[k]$ be the TLE-based orbit for $s$ propagated to epoch $k$, and
$\vb{e}_s$ be the error in $\vbt{r}_s[k]$, with
$\vb{e}_s~\sim~\mathcal{N}(\vb{0},Q_s)$.  Modeling $\vb{e}_s$ as constant is
acceptable for observation intervals limited to a few seconds as in the current
case. It is assumed that $\vb{e}_s$ and ${w}[k]$ are mutually independent for
all $s \in \mathcal{S}_\text{e}$ and $k \in \mathcal{K}$.  The true satellite location is
related to $\vbt{r}_s[k]$ and $\vb{e}_s$ by
\begin{equation}
  \label{eq:rsTRue}
  \vb{r}_{s}[k] = \vbt{r}_s[k] + \vb{e}_s, \quad s \in \mathcal{S}_\text{e}, ~~k \in \mathcal{K}
\end{equation}

The framework established in \citet{komodromos2025networkplans,
  morgan2025mockplans} can be leveraged to investigate the effects of $\vb{e}_s$
on the TDOA measurement.  Let the range Jacobian of the $i$th receiver with
respect to $\vb{r}_{s}[k]$ be given by
\begin{equation}
  \label{eq:rhoJac}
  \Gamma_i[k] = \frac{\partial \rho_{is}[k]}{\partial \vb{r}_s[k]}  = -
  \frac{\left(\vb{g}_i - \vb{r}_s[k] \right)\T}{\| \vb{g}_i - \vb{r}_s[k] \|},
  \quad  i \in \{1,2\}; ~~k \in \mathcal{K}
\end{equation}
This is approximated as constant over the short observation interval:
$\Gamma_i[k] = \Gamma_i[1], ~ i \in \{1,2\}$.  The modeled TDOA for
$s \in \mathcal{S}_\text{e}$ can be approximated by retaining up to the linear
terms in a Taylor series expansion of $\bar{y}_s[k]$ about $\vbt{r}_s[k]$:
\begin{align}
  \label{eq:ybarsModel}
  \begin{split}
    \bar{y}_s[k] &= \rho_{2s}[k] - \rho_{1s}[k] \\
                 &= \|(\vbt{r}_{s}[k] + \vb{e}_s) - \vb{g}_2 \| - \| (\vbt{r}_{s}[k] + \vb{e}_s) - \vb{g}_1 \| \\
                 &= \| \vbt{r}_{s}[k]  - \vb{g}_2 \| - \| \vbt{r}_{s}[k] - \vb{g}_1 \| + \left(\Gamma_2 - \Gamma_1 \right)\vb{e}_s  \\
                 &= \tilde{y}_{s}[k] + \eta_s
  \end{split}
\end{align}
One thus observes that the random ephemeris error $\vb{e}_s$ manifests as a
random TDOA measurement bias $\eta_s=(\Gamma_2-\Gamma_1)\vb{e}_s$ over all
$k \in \mathcal{K}$, with $\eta_s \sim \mathcal{N}(0, \sigma_s^2)$ and
$\sigma_s^2=(\Gamma_2-\Gamma_1)Q_s( \Gamma_2-\Gamma_1)\T$.  One may also lump
into $\eta_s$ any residual time synchronization error between the reference
stations, which, like $\vb{e}_s$ can be considered constant over $\mathcal{K}$.
Finally, let $\tilde{\vb{y}}_s=[\tilde{y}_s[1],\dots,\tilde{y}_s[K]]\T$ be the
TLE-based modeled TDOA vector for $s \in \mathcal{S}_\text{e}$.

\subsection{TDOA-Based Association Framework}
Based on the foregoing measurement models, an association framework can be
developed with TLE-derived satellite positions. The objective is to assess
whether a candidate satellite $s \in \mathcal{S}_\text{e}$ is consistent with
the measurement $\vb{y}$. For each $s$, the elements of $\tilde{\vb{y}}_s$ can
be readily calculated as in \eqref{eq:ybarsModel}.  Let $\vb{1}$ be the all-ones
vector of appropriate size, and let
$\vb{w} = [w[1],\dots, w[K]]\T \sim \mathcal{N}(\vb{0}, R)$, with
$R = \sigma^2 I$. Then the TDOA measurement residual with respect to the
TLE-modeled satellite $s$ is
\begin{align}
  \label{eq:tdoaResidualVector}
  \vb{\gamma}_s  = \vb{y} - \tilde{\vb{y}}_s =
  \bar{\vb{y}}_{u}  - \bar{\vb{y}}_{s} + \eta_s\vb{1} + \vb{w}
\end{align}
Note that if $s = u$, i.e., if $s$ is the true satellite, then $\vb{\gamma}_s$
is zero mean.  A cost can be assigned to each satellite $s \in \mathcal{S}_\text{e}$
under a Bayesian framework that treats $\eta_s$ as an unknown random parameter
to be estimated.  This is similar to how unknown transmitter parameters are
treated in Section \ref{sec:CNR_SatID} except that $\eta_s$ is modeled as a
random variable with a known prior distribution:
$\eta_s \sim \mathcal{N}(0, \sigma_s^2)$.  The cost function follows from the
negative log of the \emph{a posteriori} probability $p(\eta_s | \vb{y})$:
\begin{align}
  \label{eq:costFcnTDOA}
  J_s(\eta_s) = \left( \vb{\gamma}_s - \eta_s \vb{1} \right)\T R^{-1}
  \left( \vb{\gamma}_s - \eta_s\vb{1} \right) +
  \left(\frac{\eta_s}{\sigma_s} \right)^2 
\end{align}
For satellite $s$, the maximum \emph{a posteriori} estimate of $\eta_s$ is the
value that minimizes the cost: $ \eta^*_s = \argmin_{\eta_s} J_s(\eta_s)$.
Under the true hypothesis $H_u$, the minimum cost is distributed as
\begin{align}
  H_u:\quad J_u(\eta^*_u) \sim \chi^2_{K}
\end{align}
This distribution arises because the total number of independent measurements is
$K + 1$ ($K$ from $\vb{y}$ and one from the prior constraint on $\eta_u$), but
one degree of freedom is lost to estimate $\eta_u$ \citep{y_barshalom01_tan}.

Thus, as in Section \ref{subsec:implicationsForCINRSourceID}, the satellite
identification problem reduces to a multi-hypothesis association problem where
comparisons between hypotheses are made on the basis of a cost function.  In
fact, the two techniques can be combined by taking the probability
$P(s |\vb{z})$ from Section \ref{subsec:implicationsForCINRSourceID} as the
prior probability for the TDOA-based identification problem.  Then, since
$\mathcal{S}_\text{e}$ is assumed to include the true interference satellite,
and approximating $\sigma_s$ as equal for all $s \in \mathcal{S}_\text{e}$, the
probability that $H_s$ is true is
\begin{equation}
  \label{eq:probBS2}
  P(s | \vb{y},\vb{z}) = \frac{\exp\left(-\tfrac{1}{2}
      J_s(\eta^*_s) \right)  P(s | \vb{z})}
  {\displaystyle\sum_{a \in \mathcal{S}_\text{e}} \exp\left(-\tfrac{1}{2}
      J_a(\eta^*_a) \right)  P(a | \vb{z})}, \quad s \in \mathcal{S}_\text{e}
\end{equation}
Here, the vector $\vb{z}$ contains the ordered station-specific detection
statistics for those stations $\mathcal{I}_\text{d}$ that detect the
interference event associated with the TDOA measurement, i.e., values from the
set $\{\Lambda_i \mid i \in \mathcal{I}_\text{d} \}$.  As before, one can choose
the highest-probability satellite, or the top $N$ most probable, or all those
for which $P(s | \vb{y},\vb{z})$ exceeds a threshold.

\subsection{A Wideband Raw-Signal Capture} 
\begin{figure}[t]
  \centering
  \begin{minipage}[b]{0.49\textwidth}
    \centering
    \includegraphics[width=\linewidth]{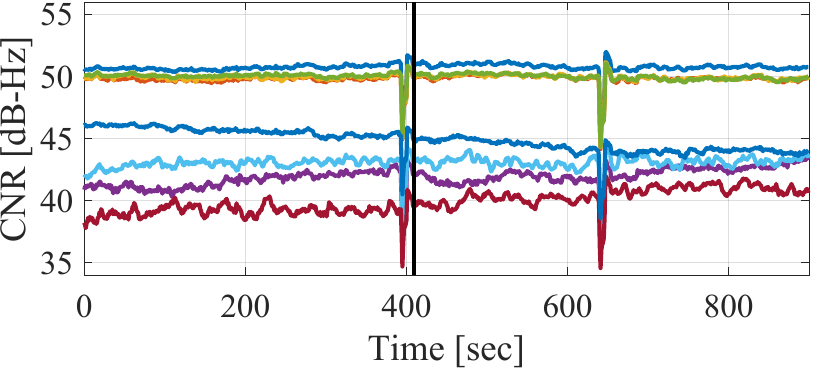}
  \end{minipage}
  \hfill
  \begin{minipage}[b]{0.49\textwidth}
    \centering
    \includegraphics[width=\linewidth]{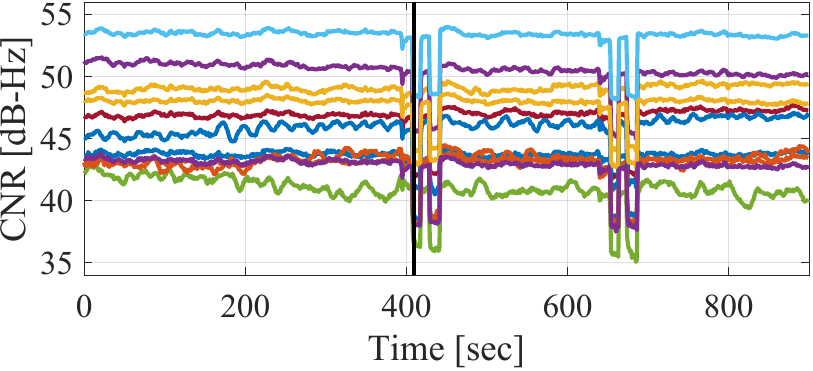}
  \end{minipage}
  \caption{CNR time history of tracked GPS L1 C/A (left) and BeiDou B1I (right)
    signals at the IGS station METG in Finland. The GPS L1 C/A signals are first
    affected by each interference event, followed by the BeiDou B1I signals.
    The black line indicates the 2.3-second time-overlapped raw IQ capture
    interval.}
  \label{fig:CNR_2026}
\end{figure}

\begin{figure}[t]
  \centering
  \begin{minipage}[b]{0.49\textwidth}
    \centering
    \includegraphics[width=\linewidth]{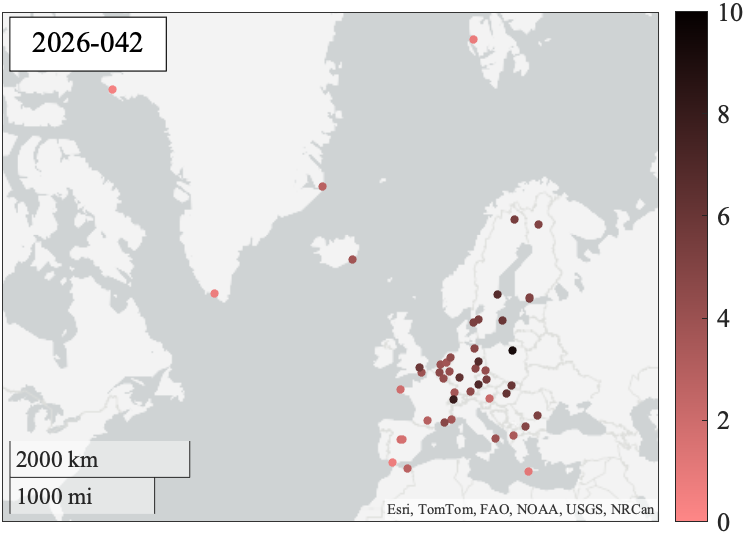}
  \end{minipage}
  \hfill
  \begin{minipage}[b]{0.49\textwidth}
    \centering
    \includegraphics[width=\linewidth]{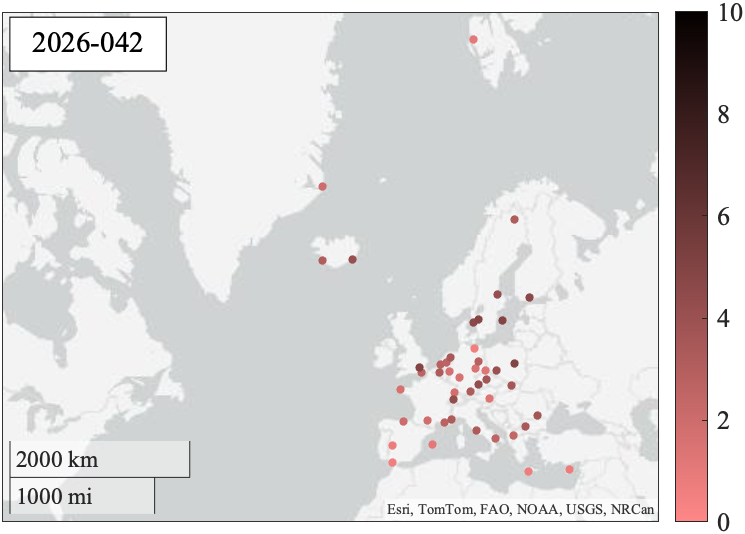}
  \end{minipage}
  \caption{Interference heat maps for day 42 of year 2026 from tracked GPS L1
    C/A (left) and BeiDou B1I (right) signals. }
  \label{fig:drop_2026}
\end{figure}

On February 11, 2026, several wide-area transient GNSS interference events
occurred across Europe.  Fig. \ref{fig:CNR_2026} shows the CNR of the tracked
GPS L1 C/A and BeiDou B1I signals at the METG station in Finland for an interval
on this day during which two events occurred.  For each event, the GPS L1 C/A
signals first dropped by 5 dB, then recovered, immediately following which the
BeiDou B1I signals experienced two cycles of 5 dB degradation and recovery.  The
CNR drops of the BeiDou B1I signals persisted for ten seconds---more than twice
as long as the drops observed on the GPS L1 C/A signals.  The interference heat
maps shown in Fig.~\ref{fig:drop_2026} reveal a spatial pattern consistent with
previously observed interference events.

The remainder of this section analyzes contemporaneous raw IQ samples captured
by a receiver located in Amsterdam, Netherlands (R1), and one located in
Trondheim, Norway (R2).  R1 employed a complex sampling rate of 60~MHz, while R2
sampled at 75~MHz. Both captures were centered at 1585~MHz, utilized three-bit
quantization, and were driven by high-stability oven-controlled crystal
oscillators (OCXOs). These receivers are part of the Advanced RFI Detection,
Analysis, and Alerting System, which focuses on the capture and collection of
GNSS interference signals \citep{sokolova2022characterization,
  Morrisonnavi.560}.  The system was developed under the European Space Agency
Navigation, Innovation, and Support Program (NAVISP).

R1 captured the interference signals responsible for the spatial
degradation pattern shown in Fig.~\ref{fig:drop_2026} and the power spectrum
shown in Fig.~\ref{fig:PSD}.  The interference waveform that caused CNR drops on
the GPS~L1~C/A signals was centered at 1577.5~MHz, which is consistent with the
interference observed in Fig.~\ref{fig:spectrum}.  This signal exhibited
cyclostationary properties, with repeating structure every 12~us.  The
interference waveform that caused the CNR drops on the BeiDou B1I signals was
centered at 1558.5~MHz. It too exhibited cyclostationary properties, at first
with repeating structure every 255.8~us, then later at 292.9~us.

On February 11, 2026 at 05:36:30 UTC, R1 and R2 recorded 2.3~seconds of
time-overlapped raw IQ samples during the 1558.5-MHz-centered interference
event.  Several GPS and Galileo signals were contained within the captured
bandwidth, which permitted processing and GPS time registration to better than
30~ns using the GRID software-defined GNSS receiver
\citep{nichols2022launch,pany2024historysdr}.

\subsection{TDOA Measurement and Method Validation}
\begin{figure}[t]
  \centering
  \begin{minipage}[b]{0.49\textwidth}
    \centering
    \includegraphics[width=\linewidth]{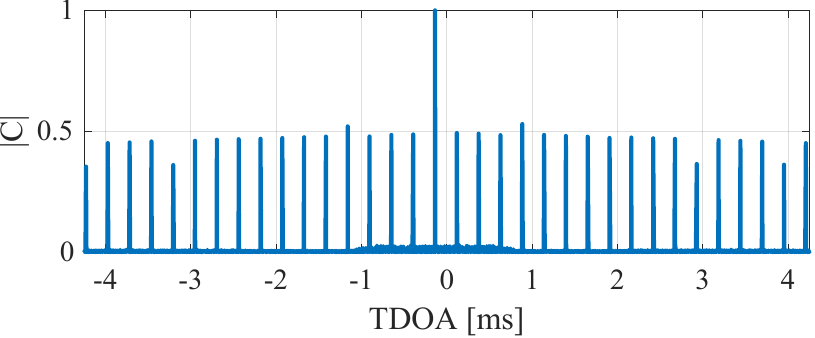}
  \end{minipage}
  \hfill
  \begin{minipage}[b]{0.49\textwidth}
    \centering
    \includegraphics[width=\linewidth]{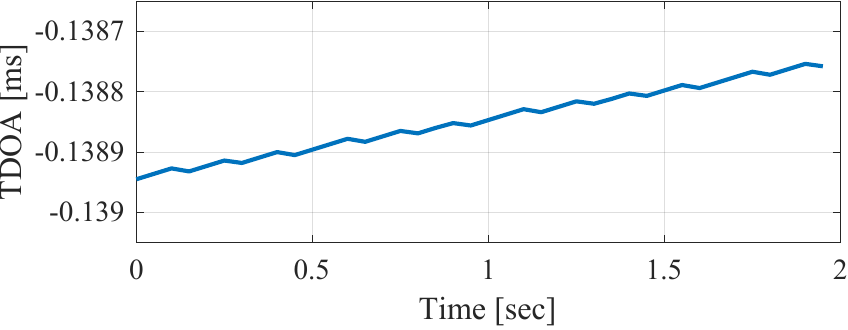}
  \end{minipage}
  \caption{Left: Example normalized cross correlation at the optimal
    FDOA. Right: TDOA measurement time history.}
  \label{fig:correlation}
\end{figure}

The IQ samples from the two receivers were synchronized in both time and
frequency and the samples from R1 were upsampled to 75-MHz to match those of R2.
To obtain the maximum likelihood estimates of TDOA and frequency-difference of
arrival (FDOA), the complex ambiguity function (CAF) was computed
\citep{stein81_afp}.  The TDOA and FDOA values that maximize the CAF magnitude
correspond to the maximum likelihood estimates.  The left panel of
Fig.~\ref{fig:correlation} shows the magnitude of the normalized cross
correlation $C$ at the optimal FDOA estimate (155~Hz) for a 50~ms integration
interval. Similar functions were repeatedly calculated every 50 ms over two
seconds to obtain the 40 TDOA measurements shown as a time history in the right
panel of Fig.~\ref{fig:correlation}.  The TDOA measurements are
magnitude-bounded by 4.25~ms, which corresponds to the delay of a hypothetical
light pulse traveling between the two receivers.  The first 0.3~seconds of the
time-overlapped IQ samples were excluded because near-equal-magnitude
cyclostationary peaks every 255.8~us caused ambiguity in the CAF.  Apparently,
the interference signal was purely periodic during this interval.  After 0.3
seconds, significant repetition remained present, but the CAF peak became
unambiguous, as shown in the left panel of Fig. \ref{fig:correlation}.

\begin{figure}[t]
  \centering
  \begin{minipage}[b]{0.49\textwidth}
    \centering
    \includegraphics[width=\linewidth]{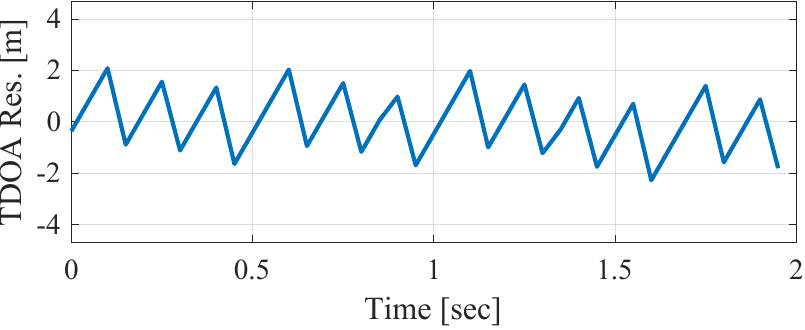}
  \end{minipage}
  \hfill
  \begin{minipage}[b]{0.49\textwidth}
    \centering
    \includegraphics[width=\linewidth]{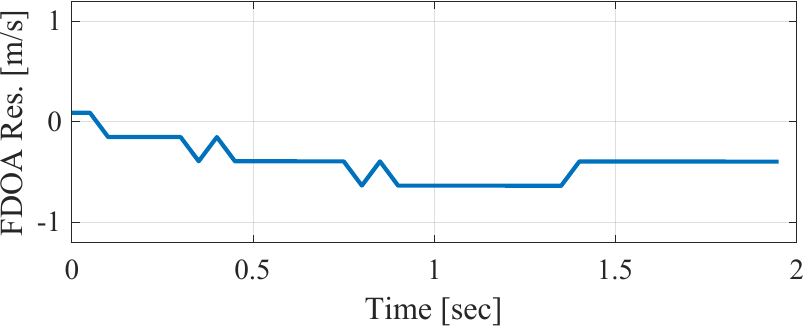}
  \end{minipage}
  \caption{TDOA (left) and FDOA (right) residuals for Cosmos 2546 (NORAD ID
    45608).  The effects of nearest-sample quantization and Doppler search
    quantization are evident in the respective plots.}
  \label{fig:residuals}
\end{figure}
To validate the TDOA measurement technique, TDOAs were measured for each
available GPS L1 C/A signal captured in the wideband data and compared against
the expected TDOA for the corresponding overhead GPS satellite. Expected TDOA
values were modeled using the true satellite positions as provided by IGS final
ephemerides.  Besides differential time-of-flight, the model also accounted for
tropospheric delay via the Saastamoinen model, ionospheric delay via the
Klobuchar model, and ECEF frame rotation during signal time of flight. The
resulting measured and modeled TDOAs exhibited excellent agreement, remaining
within 15~ns of each other during the $2.3$-second overlapping interval of the
wideband captures.

\subsection{Satellite Association}
The measurement vector $\vb{y}$ was formed from the time history shown in
Fig. \ref{fig:correlation}.  Values for the TDOA measurement noise variance
$\sigma^2$ and for the TLE-based position error variance $Q_s$ were chosen as
follows for all $s \in \mathcal{S}_\text{e}$.  Because the interference signal
has a high interference-to-noise ratio and a bandwidth spanning several MHz, the
TDOA estimation errors are dominated by nearest-sample quantization error, the
variance of which, denoted $\sigma_\text{n}^2$, can be determined by calculating
the variance of a uniform distribution with a width of one sample, or
$\sigma_\text{n}^2=0.76^2~\text{m}^2$ for 75-MHz sampling.  Alternatively, a
conservative choice of $\sigma^2$ can be determined by selecting the width of
half a sample, $\sigma^2=2^2~\text{m}^2$ for 75-MHz sampling, which is an
over-bound when compared to $\sigma_\text{n}^2$ and more amenable to the
developed Gaussian framework. Thus, $\sigma^2 = 2^2$~m.

Position estimates derived from propagated TLEs typically exhibit
kilometer-level inaccuracies, which can escalate to 10~km under volatile orbital
conditions \citep{vetter2007fifty}.  Error is typically concentrated in the
along-track direction and increases with the age of ephemeris.  A conservative
approach was taken to over-bound the ephemeris position error.  For all
$s \in \mathcal{S}_\text{e}$, the position covariance in ECEF was set to
$Q_s=10000^2I$.  The TLEs of satellites identified as either ``payload" or
``unknown" within seven days of the interference event were retrieved from
space-track.org. For each satellite, the TLE with the epoch nearest to the
measurement time was used to calculate position.  The set $\mathcal{S}_\text{e}$
was composed of all satellites satisfying elevation angle masking with
$\alpha_0 = 0$ for all stations that detected the event.

For each satellite $s \in \mathcal{S}_\text{e}$, the cost $J_s(\eta^*_s)$ was
computed.  For a threshold $\nu$ chosen such that
$P(J_u(\eta^*_u) > \nu) \leq 10^{-3}$, only one satellite satisfied
$J_s(\eta^*_s) \leq \nu$: Cosmos~2546~(NORAD~ID~45608).  This is one of six
Molniya-orbit satellites that compose the Russian Edinaya Kosmicheskaya Sistema
(EKS), an early warning constellation. For Cosmos~2546, even assuming equally
probable priors $P(s | \vb{z}) = 1/|\mathcal{S}_\text{e}|$ for
$s \in \mathcal{S}_\text{e}$, the probability $P(s|\vb{y}, \vb{z})$ calculated
according to \eqref{eq:probBS2} for Cosmos~2546 was numerically
indistinguishable from unity.  Cosmos~2546 also satisfied an
$\alpha_0 = 35^\circ$ elevation mask with respect to every
$i \in \mathcal{I}_\text{d}$.  This is consistent with its highly elliptical
orbit, which positioned it far above the northern Atlantic Ocean during the
event captured by the wideband data.

T/FDOA residuals for Cosmos~2546 are shown in Fig.~\ref{fig:residuals}.  The
estimated value of the TLE error was $\eta^*_s = 188$~m and the standard
deviation of the TDOA residuals was $1.2$~m, indicating a highly plausible
match.  The projected TDOA standard deviation due to TLE error was $\sigma_s=370$
m for this satellite given the assumed $Q_s$.  The FDOA residuals shown in the
right panel of Fig. \ref{fig:residuals} also indicate tight consistency.  FDOA
measurements were not used for association but could easily be incorporated into
an augmented association framework.

\subsection{Further Analysis and Discussion}
When all space objects are considered, not only those in $\mathcal{S}_\text{e}$,
one additional satellite satisfies $J_s(\eta^*_s) < \nu$---one from SpaceX's
Starlink constellation.  But this
satellite is not a viable candidate: during the interval of the TDOA
measurement it was located over the Pacific Ocean west of South America,
yielding elevation angles far below zero for all stations detecting the event.
One may conclude from these false associations that fusing TDOA-based
association with elevation angle considerations or CNR-based priors is necessary
for a confident unique association when data are limited to a single short TDOA
time history.

The TLE for Cosmos~2546 closest to the TDOA measurement time is 40 hours offset,
which motivates further examination of its expected accuracy.  An analysis was
performed based on the full historical archive of TLEs for Cosmos~2546 since its
launch in 2020.  Let $\mathcal{L} = \{1,\dots, L\}$ be the index set for this
TLE archive, and let $t_l$ represent the epoch for the $l$th TLE.  Let
$\vb{r}_k(t_l)$ denote the position of Cosmos~2546 based on the $k$th TLE but
propagated to the epoch of the $l$th TLE.  For all $l \in \mathcal{L}$ and all
$k \in \mathcal{L} \setminus l$, one may assume that $\vb{r}_l(t_l)$ is more
accurate than $\vb{r}_k(t_l)$.  Let $\mathcal{L}_l \subset \mathcal{L}$ be the
set of TLEs within a 60-hour window of $t_l$, and let
$\mathcal{R}_l = \{ \|r_k(t_l) - r_l(t_l)\| \mid k \in \mathcal{L}_l \}$ be the
corresponding residual set.  Statistical analysis of
$\mathcal{R} = \bigcup_{l \in \mathcal{L}} \mathcal{R}_l$ (after outlier
removal) revealed that the maximum per-dimension residual standard deviation was
below 4~km and occurred in the along-track direction. This value represents an
appropriate upper bound for Cosmos~2546's \emph{a priori} position error and
falls below 10~km, the determinant of $Q_s$ assumed in the association
framework.  This methodology for gauging orbital accuracy mirrors established
techniques for verifying the accuracy of the IGS's real-time service products
\citep{hadas2015igs, griffiths2009precision}.

Strictly speaking, identification of Cosmos~2546 as a source for the transient
wide-area interference holds only for the 1558.5~MHz interference during the
time of wideband IQ capture.  But due to the time proximity with the 1577.5~MHz
interference, and to their similar time-domain and spatial patterns,
interference in the two bands around the time of the IQ capture almost surely
originated from the same satellite.

Two other assumptions should be acknowledged as underlying this paper's results:
(1) the TLEs for all interference sources are available on space-track.org, and
(2) neither Cosmos~2546 nor any satellite near the quasi-hyperboloid surface
defined by the TDOA measurement time history in Fig. \ref{fig:correlation}
maneuvered so significantly between its nearest TLE epoch and the TDOA
measurement as to alter the association findings.  Both assumptions seem
reasonable.  In fact, it was found that the TLE error standard deviation in
$Q_s$ could be expanded from $10$~km to $500$~km before any additional
satellites became consistent with the TDOA data.

Note that Cosmos~2546 was launched in May 2020 and so cannot be responsible for
the interference events that occurred in 2019.  Moreover, Cosmos~2546 was not
over Europe during some interference events after May 2020.  But during all
events on the 75 days shown in Table \ref{tbl:occurances} there was at least
one EKS satellite above a 35$^\circ$ elevation angle with respect to every
reference station that observed the interference.  Thus, it is highly probable
that the EKS constellation is collectively responsible for the wide-area
transient GNSS interference events noted since 2019.

\section{Conclusion} This paper presented a comprehensive analysis of a novel
GNSS interference phenomenon: wide-area transient interference from a
space-based source causing up to 10-dB GNSS degradation across Europe since 2019
in the important L1 band.  The interference's spatial, temporal, and spectral
properties were detailed.  A framework was developed to detect events using 1-Hz
carrier-to-noise ratio observables from a network of 165 reference stations.  A
total of 75~days were identified since 2019 on which at least one wide-area
interference event occurred.  Three techniques were developed to identify
candidate source satellites: (1) a simple technique that finds all
satellites satisfying a given elevation mask (e.g., 0$^\circ$) for each station
detecting the interference; (2) a more advanced technique based on each
station's detection statistic rather than on a binary decision, and on
estimation of unknown parameters such as the interference transmit power; and
(3) a technique based on time-difference-of-arrival measurements made possible
by capture of wideband (e.g., 60-MHz) raw samples during an interference event.
By a combination of these techniques the satellite Cosmos~2546~(NORAD~ID~45608)
was identified with high confidence as one source of the interference.  Further
analysis pointed to the Russian Edinaya Kosmicheskaya Sistema, an early warning
constellation to which Cosmos~2546 belongs, as collectively responsible for the
wide-area transient interference causing GNSS degradation across Europe since
2019.

\section*{Acknowledgments} Research was supported by the U.S. Department of
Transportation under Grant 69A3552348327 for the CARMEN+ University
Transportation Center, and by affiliates of the 6G@UT center within the Wireless
Networking and Communications Group at The University of Texas at Austin.
Special thanks to Aiden Morrison from SINTEF Digital for providing dual-station
wideband-sampled data, and to Michael Meurer and Steffen Th\"{o}lert from the
German Aerospace Center for calibrated estimates of GPS transmitter power.
